\def\thisisTR{}

\newif\iftr
\ifdefined\thisisTR\trtrue\else\trfalse\fi
\newif\ifpaper
\iftr\paperfalse\else\papertrue\fi

\iftr
\documentclass[a4paper,10pt]{article}
\usepackage{styles/arxiv}
\else
\documentclass[10pt]{styles/llncs}
\fi

\usepackage{etoolbox}

\newtoggle{tr}
\iftr
\toggletrue{tr}
\else
\togglefalse{tr}
\fi

\usepackage[multidot]{grffile}
\usepackage{xspace}

\usepackage[activate={true,nocompatibility},final,tracking=true,kerning=true,spacing=true,stretch=10,shrink=30]{microtype}
\microtypecontext{spacing=nonfrench}

\usepackage[utf8]{inputenc}
\usepackage{flushend}

\usepackage{amsmath,amssymb,amsfonts,bm}

\usepackage[english]{babel}
\usepackage{csquotes}
\usepackage[inline,shortlabels]{enumitem}
\usepackage{subcaption}
\usepackage{overpic}
\usepackage{booktabs}
\usepackage{tablefootnote}
\newcommand*{\thead}[1]{\multicolumn{1}{c}{\bfseries #1}}
\newcommand*{\tlhead}[1]{\multicolumn{1}{l}{\bfseries #1}}

\usepackage{pifont}
\usepackage{nicefrac}
\usepackage[noend]{algpseudocode}
\usepackage{algorithm}
\algnewcommand{\Label}[1]{\Statex \hspace{-\algorithmicindent}\hspace{-\labelsep} \textbf{#1}}
\algnewcommand{\Input}{\item[\rlap{\textbf{Input:}}{\hphantom{\textbf{Output:}}}]}
\algnewcommand{\Output}{\item[\textbf{Output:}]}
\algnewcommand{\Seq}{\item[\phantom{\textbf{Output:}}]}
\algnewcommand{\IIf}[1]{\State\algorithmicif\ #1\ \algorithmicthen}
\algnewcommand{\EndIIf}{\unskip\ \algorithmicend\ \algorithmicif}
\newcommand{\cupEq}{\,\,\cup\!\!=}

\DeclareMathOperator*{\argmin}{arg\,min}

\algnewcommand\algorithmicparfor{\textbf{parfor}}
\algblockdefx{ParFor}{EndParFor}[1]
  {\algorithmicparfor\ #1\ \algorithmicdo}
  {\algorithmicend\ \algorithmicparfor}

\makeatletter
\ifthenelse{\equal{\ALG@noend}{t}}%
  {\algtext*{EndParFor}}
  {}%
\makeatother

\makeatletter
\def\@xfootnote[#1]{%
  \protected@xdef\@thefnmark{#1}%
  \@footnotemark\@footnotetext}
\makeatother

\usepackage{tikz}
\usetikzlibrary{calc}

\usepackage[numbers,sectionbib,sort&compress]{natbib}
\makeatletter
\def\NAT@spacechar{~}
\makeatother

\usepackage{siunitx}
\usepackage{placeins}

\usepackage{hyperref}

\newcommand {\ie}{\mbox{i.\,e.}\xspace}     
\newcommand {\eg}{\mbox{e.\,g.}\xspace}     
\newcommand{\dac}{D\&C\xspace}

\newcommand{\tabref}[1]{Table\;\ref{#1}\xspace}
\newcommand{\figref}[1]{Figure\;\ref{#1}\xspace}
\newcommand{\secref}[1]{Section\;\ref{#1}\xspace}

\renewcommand{\algref}[1]{Algorithm\;\ref{#1}\xspace}
\newcommand{\lnref}[1]{Line\;\ref{#1}\xspace}

\makeatletter
\newcommand{\eqnref}{\@ifstar
			\eqnrefS%
			\eqnrefNS%
		    }
\makeatother
\newcommand{\eqnrefNS}[1]{Equation\;\eqref{#1}\xspace}
\newcommand{\eqnrefS}[1]{Equation\;\ref{#1}\xspace}

\newcommand{\set}[1]{\ensuremath{\mathbf{#1}}}

\newcommand{\vertices}[2][]{%
  \ifstrempty{#1}{%
      \ensuremath{\operatorname{vertices}(#2)}
    }{%
      \ensuremath{\operatorname{vertices}_{#1}(#2)}
    }%
}

\newcommand{\neighbors}[2][]{%
  \ifstrempty{#1}{%
      \ensuremath{\operatorname{neighbors}(#2)}
    }{%
      \ensuremath{\operatorname{neighbors}_{#1}(#2)}
    }%
}

\usepackage{etoolbox}

\title{Load-Balancing for Parallel Delaunay Triangulations}

\iftoggle{tr}{
\author{
  Daniel Funke\\
  Karlsruhe Institute of Technology\\
  Karlsruhe, Germany\\
  \texttt{funke@kit.edu} \\
   \And
  Peter Sanders\\
  Karlsruhe Institute of Technology\\
  Karlsruhe, Germany\\
  \texttt{sanders@kit.edu} \\
  \And
  Vincent Winkler\\
  Karlsruhe Institute of Technology\\
  Karlsruhe, Germany\\
  \texttt{vincent.winkler@student.kit.edu} \\
}
}{
\author{Daniel Funke \and Peter Sanders \and Vincent Winkler}
\institute{Karlsruhe Institute of Technolgy, Karlsruhe, Germany\\
    \email{\{funke, sanders\}@kit.edu, vincent.winkler@student.kit.edu}}
}

\begin{document}
\maketitle

\begin{abstract}
Computing the Delaunay triangulation (DT) of a given point set in $\mathbb{R}^D$
is one of the fundamental operations in computational geometry.
Recently, \citet{Funke2017} presented a divide-and-conquer DT algorithm that
merges two partial triangulations
by re-triangulating a small subset of their vertices
-- the \emph{border} vertices --
and combining the three triangulations efficiently via parallel hash table lookups.
The input point division should therefore yield
roughly equal-sized partitions for good load-balancing
and also result in a small number of border vertices for fast merging.
In this paper, we present a novel divide-step based on partitioning 
the triangulation of a small sample of the input points.
In experiments on synthetic and real-world data sets, we achieve
nearly perfectly balanced partitions and small border triangulations.
This almost cuts running time in half compared to non-data-sensitive division
schemes on inputs exhibiting an exploitable underlying structure.
\end{abstract}

\section{Introduction}\label{sec:intro}

The Delaunay triangulation (DT) of a given point set in $\mathbb{R}^D$ has
numerous applications in computer graphics, data visualization, 
terrain modeling, pattern recognition and finite element methods~\citep{Kohout2005}.
Computing the DT is thus one of the fundamental operations in geometric computing.
Therefore, many algorithms for efficiently computing the DT have been proposed (see survey in~\citep{Su1995})
and well implemented codes exist \citep{cgal:hs-chdt3-15b,Shewchuk1996}.
With ever increasing input sizes, research interest has shifted from sequential algorithms 
towards parallel ones\iftoggle{tr}{ \citep{Batista2010,Blelloch1999,Cignoni1998,Kohout2005,Fuetterling2014,Chen2012}}{}.

Recently, we presented a novel divide-and-conquer (\dac) DT algorithm for arbitrary dimension~\citep{Funke2017}
that lends itself equally well to shared and distributed memory parallelism
and thus hybrid parallelization.
While previous \dac DT algorithms suffer from a complex -- often sequential --
divide or merge step~\citep{Cignoni1998,Lee2001},
our algorithm reduces the merging of two partial triangulations to re-triangulating a small subset of
their vertices -- the \emph{border} vertices -- using the same parallel algorithm
and combining the three triangulations efficiently via hash table lookups.
All steps required for the merging
-- identification of relevant vertices, triangulation and combining the partial DTs --
are performed in parallel.

The division of the input points in the divide-step needs to address a twofold
sensitivity to the point distribution:
the partitions need to be approximately equal-sized for good load-balancing,
while the number of \emph{border vertices} needs to be minimized for fast merging.
This requires partitions that have many internal Delaunay edges but only few
external ones, \ie a graph partitioning of the DT graph.
In this paper we propose a novel divide-step that approximates this graph partitioning
by triangulating and partitioning a small sample of the input points,
and divides the input point set accordingly.

The paper is structured as follows:
we review the problem definition, related work on partitioning for DT algorithms
and our \dac DT algorithm from~\citep{Funke2017} in \secref{sec:prelim}. 
Subsequently, our proposed divide-step is described in \secref{sec:sbp},
along with a description of fast intersection tests
for the more complexly shaped partition borders and implementation notes.
We evaluate our algorithms in \secref{sec:eval} 
and close the paper with conclusions and an outlook on future work in \secref{sec:outro}.

\section{Preliminaries}\label{sec:prelim}

\subsection{Delaunay Triangulations}\label{sec:def}

\iftoggle{tr}{%
A $d$-simplex is a generalization of a triangle ($d = 2$) to $d$-dimensional space.
A $d$-simplex $s$ is a $d$-dimensional polytope, \ie~the convex hull of $d+1$ points.
The convex hull of $m+1$ of these $d+1$ points is called an $m$-face of $s$.
Specifically, the $0$-faces are the vertices of $s$
and the $(d-1)$-faces are its facets.}{}
Given a $d$-dimensional point set $\set{P} = \{p_1, p_2, \dots, p_n\} \subset \mathbb{R}^d$ for all $i \in \{ 1, \dots, n \}$,
a triangulation $T(\set{P})$ is a subdivision of the convex hull of $\set{P}$ 
into $d$-simplices such that the set of vertices of $T(\set{P})$ coincides with $\set{P}$
and any two simplices of $T$ intersect in a common $d-1$ facet or not at all.
The union of all simplices in $T(\set{P})$ is the convex hull of point set $\set{P}$.
A Delaunay triangulation $DT(\set{P})$ is a triangulation of $\set{P}$ such that
no point of $\set{P}$ is inside the circumhypersphere of any simplex in $DT(\set{P})$.
\iftoggle{tr}{%
If the points of $\set{P}$ are in \emph{general position}, 
\ie~no $d+2$ points lie on a common $d$-hypersphere,
$DT(\set{P})$ is unique~\citep{Delaunay1934}.}{} 

\subsection{Related Work}\label{sec:rw}

Many algorithms for the parallel construction of the DT of a given point set
have been proposed in the literature.
They generally fall into one of two categories:
parallel incremental insertion and \dac approaches.
We will focus on a review of the divide-step of the latter.
A more comprehensive discussion of both algorithm types is given in
\citep{Funke2017}.

\citet{Aggarwal1988} propose the first parallel \dac DT algorithm.
They partition the input points along a vertical line into blocks,
which are triangulated in parallel and then merged sequentially.
The authors  do not prescribe how to determine the location of the splitting line.
\citet{Cignoni1998} partition the input along cutting (hyper)planes and
firstly construct the simplices of the triangulation crossing those planes
before recursing on the two partitions.
The remaining simplices can be created in parallel in the divided regions
without further merging.
The authors mention that the regions should be of roughly equal cardinality,
but do not go into the details of the partitioning.
\citet{Chen2010} and~\citet{Lee2001} explicitly require splitting along the
median of the input points.
Whereas the former uses classical splitting planes,
the latter traces the splitting line with Delaunay edges,
thus eliminating the need for later merging.

The subject of input partitioning has received more attention in the meshing community.
A \emph{mesh} of a point set $\set{P}$ is a triangulation of every point in
$\set{P}$ and possibly more -- so called \emph{Steiner points} -- to refine the
triangulation\iftoggle{tr}{~\citep{Cheng2012}}{}.
\citet{Chrisochoides2006} surveys algorithms for parallel mesh generation and
differentiates between continuous domain decomposition
-- using quad- or oct-trees --
and discrete domain decomposition using an initial coarse mesh that is
partitioned into submeshes, trying to minimize the surface-to-volume ratio of
the submeshes.
\citet{Chrisochoides2000} propose an algorithm that meshes the
subproblems via incremental insertion using the Bowyer-Watson algorithm.

\subsection{Parallel Divide-and-Conquer DT Algorithm}\label{sec:algo}

\iftoggle{tr}{
\begin{algorithm*}[tbph]
\begin{algorithmic}[1]
\Input points $\set{P} = \{p_1, \dots, p_n \}$ with $p_i \in \mathbb{R}^D$
\Output Delaunay triangulation $DT(\set{P})$

\If{$n < N$}
\State \Return $\operatorname{\emph{sequentialDelaunay}}(\set{P})$
  \Comment base case
\EndIf
\State  $\begin{pmatrix}
         \set{P}_1 & \dots & \set{P}_k
        \end{pmatrix} 
        \gets \operatorname{partitionPoints}(\set{P}, k)$
          \Comment partition points into $k$ partitions
\State  $\begin{pmatrix}
         T_1 & \dots & T_k
        \end{pmatrix}
        \gets \begin{pmatrix}
        \operatorname{Delaunay}(\set{P}_1) & \dots & \operatorname{Delaunay}(\set{P}_k)
        \end{pmatrix}$ \Comment \emph{parallel} triangulation

\smallskip
\Label{Border triangulation:}
\State  $\set{B} \gets \varnothing; \quad
         \set{Q} \gets \bigcup_{1 \leq i \leq k} \operatorname{convexHull}(T_i)$
  \Comment initialize set of border simplices
\ParFor{$s_{i, x} \in \set{Q}$} \Comment simplex originating from triangulation $T_i$
  \State $\operatorname{mark}(s_{i, x})$ \Comment process each simplex only once
  \If{$\operatorname{intersects}\left( \operatorname{circumsphere}(s_{i,x}),
      T_{j} \right) \text{, with $i \neq j$}$} \label{ln:intersectionTest}
    \State $\set{B} \cupEq \{ s_{i,x} \}$ \Comment circumsphere intersects other partition, $s_{i,x}$ is border simplex
    \For{$s_{i,y} \in \neighbors{s_{i,x}} \wedge  \neg \operatorname{marked}(s_{i, y})$} \Comment process all neighbors
      \State $\set{Q} \cupEq s_{i,y}; \quad \operatorname{mark}(s_{i, y})$
    \EndFor
  \EndIf
\EndParFor
\State $T_B \gets \operatorname{Delaunay}(\vertices{\set{B}})$
       \Comment triangulate points of border simplices
       \label{ln:borderpoints}

\smallskip
\Label{Merging:}
\State $T \gets \left( \bigcup_{1 \leq i \leq k} T_k \right) \setminus \set{B}; \quad \set{Q} \gets \varnothing$ 
  \Comment merge partial triangulations, strip border
\ParFor{$s_b \in T_B$} \Comment merge simplices from border triangulation
  \If{$\vertices{s_b} \not\subset \set{P}_i \quad \forall 1 \leq i \leq k$}
    \State $T \cupEq \{ s_b \}; \quad Q \cupEq \{ s_b \}$
      \Comment $s_b$ spans multiple partitions
  \Else
    \If{$\exists s \in \set{B}: \vertices{s} = \vertices{s_b}$}
      \State $T \cupEq \{ s_b \}; \quad Q \cupEq \{ s_b \}$ \Comment $s_b$ replaces border simplex
    \EndIf
  \EndIf
\EndParFor

\smallskip
\Label{Neighborhood update:}
\ParFor{$s_x \in Q$} \Comment update neighbors of inserted simplices
  \For{$d \in \{ 1, \dots, D+1 \}$}
    \If{$\neighbors[d]{s_x} \not\in T$} \Comment neighbor no longer in triangulation
      \State $C \gets \{s_c: f_d(s_x) = f_d(s_c) \}$ \Comment candidates with same facet hash
      \For{$s_c \in C$}
        \If{$| \vertices{s_x} \cap \vertices{s_c}| = D$ 
          }
          \State $\neighbors[d]{s_x} \gets s_c; \quad Q \cupEq s_c$ \Comment $s_c$ is neighbor of $s_x$
        \EndIf
      \EndFor
    \EndIf
  \EndFor
\EndParFor
\State \Return $T$
\end{algorithmic}
\caption{$\operatorname{Delaunay}(\set{P})$: shared memory parallel \dac algorithm}
\label{alg:sma}
\end{algorithm*}%
}

Recently, we presented a parallel divide-and-conquer algorithm for
computing the DT of a given point set~\citep{Funke2017}.
Our algorithm recursively divides the input into two partitions
which are triangulated in parallel.
The contribution lies in a novel merging step for the two partial triangulations
which re-triangulates a small subset of their vertices and combines the three
triangulations via parallel hash table lookups.
For each partial triangulation the \emph{border} is determined,
\ie the simplices whose circumhypersphere intersects the bounding box of the
other triangulation.
The vertices of those border simplices are then re-triangulated to obtain the
border triangulation.
The merging proceeds by combining the two partial triangulations,
stripping the original border simplices and
adding simplices from the border triangulation iff
\begin{enumerate*}[i)]
  \item they span multiple partitions; or
  \item are contained within one partition but exist in the same form in the
    original triangulation.
\end{enumerate*}
\iftoggle{tr}{%
We adapt the original algorithm to an arbitrary number of partitions
in \algref{alg:sma}.}{}

The algorithm's sensitivity to the input point distribution is twofold:
the partitions need to be of equal size for good load-balancing between the available cores and
the number of simplices in the border needs to minimized in order to reduce merging overhead.
As presented in~\citep{Funke2017}, the algorithm splits the input into two partitions along a hyperplane.
Three strategies to choose the splitting dimension are proposed: 
\begin{enumerate*}[i)]
 \item constant, predetermined splitting dimension;
 \item cyclic choice of the splitting dimension -- similar to $k$-D trees\iftoggle{tr}{~\citep{Bentley1975}}{}; or
 \item dimension with largest extend.
\end{enumerate*}
This can lead to imbalance in the presence of non-homogeneously structured inputs,
motivating the need for more sophisticated partitioning schemes.

\section{Sample-based Partitioning}\label{sec:sbp}

\iftoggle{tr}{
\begin{algorithm*}[tbp]
\begin{algorithmic}[1]
\Input points $\set{P} = \{p_1, \dots, p_n \}$ with $p_i \in \mathbb{R}^D$,
number of partitions $k$
\Output partitioning $\begin{pmatrix} \set{P}_1 & \dots & \set{P}_k \end{pmatrix}$

\State $\set{P}_S \gets \text{choose $\eta(n)$ from $\set{P}$ uniformly at random}$
  \Comment $\eta(n)$ sample size
\State $T \gets \operatorname{Delaunay}(\set{P}_S)$
\State $G = (V, E, \omega)$ with $V = \set{P}_S$, $E = T$ and weight function $\omega$
\State $\begin{pmatrix} V_1 & \dots & V_k \end{pmatrix} \gets \operatorname{partition}(G)$
  \Comment partition graph
\State $\begin{pmatrix} \set{P}_1 & \dots & \set{P}_k \end{pmatrix}
  \gets \begin{pmatrix} \varnothing & \dots & \varnothing \end{pmatrix}$
\ParFor{$p \in \set{P}$}
\State $v_n \gets \argmin_{v \in \set{P}_S} ||p - v||$
\Comment find nearest sample point to $p$
\State $\set{P}_i \cupEq p \text{ with } i \in [1 \dots k]: v_n \in V_i$
\Comment assign $p$ to $v_n$'s partition \label{ln:partition:nn}
\EndParFor
\State \Return $\begin{pmatrix} \set{P}_1 & \dots & \set{P}_k \end{pmatrix}$

\end{algorithmic}
\caption{$\operatorname{partitionPoints}(\set{P}, k)$: partition input into $k$ partitions.}
\label{alg:partition}
\end{algorithm*}
}

In this paper, we propose more advanced strategies for partitioning the input points than
originally presented in~\citep{Funke2017}.
The desired partitioning addresses both data sensitivities of
\iftoggle{tr}{\algref{alg:sma}}{our algorithm}.
The underlying idea is derived from sample sort\iftoggle{tr}{~\citep{samplesort}}{}:
gain insight into the input distribution from a (small) sample of the input.
\iftoggle{tr}{\algref{alg:partition} describes our partitioning procedure.}{}
A sample $\set{P}_S$ of $\eta(n)$ points is taken from the input point set of
size $n$ and triangulated to obtain $DT(\set{P}_S)$.
A similar approach can be found in Delaunay hierarchies,
were the sample triangulation is used to speed up point location queries
\citep{dt_hierarchy}.

Instead, we transform the DT into a graph $G = (V, E, \omega)$,
with $V$ being equal to the sample point set $\set{P}_S$
and $E$ containing all edges of $DT(\set{P}_S)$.
The resulting graph is then partitioned using a graph partitioning tool
to obtain a partition into $k$ blocks.

The choice of weight function $\omega$ influences the quality of the
resulting partitioning.
As mentioned in \secref{sec:algo}, the \dac algorithm is sensitive to the
balance of the blocks as well as the size of the border triangulation.
The former is ensured by the imbalance parameter $\epsilon$ of the graph
partitioning, which guarantees that for all partitions $i$:
$\left|V_i\right| \leq (1+\epsilon) \lceil  \frac{|V|}{k} \rceil$.
The latter needs to be addressed by the edge weight function $\omega$ of the graph.
In order to minimize the size of the border triangulation,
dense regions of the input points should not be cut by the partitioning.
Sparse regions of the input points result in long Delaunay edges in the sample
triangulation.
As graph partitioning tries to minimize the weight of the cut edges,
edge weights need to be inversely related to the Euclidean length of the
edge.
\iftoggle{tr}%
{\tabref{tab:weights} provides an overview of the edge weight functions considered,
which are evaluated in \secref{sec:eval:params}.}%
{In \secref{sec:eval:params} we evaluate several suitable edge weight functions.}

Given the partitions of the sample vertices $(V_1 \dots V_k)$,
the partitioning needs to be extended to the entire input point set.
The dual of the Delaunay triangulation of the sample point set
-- its Voronoi diagram -- defines a partitioning of the Euclidean space
$\mathbb{R}^d$ in the
following sense:
each point $p_{S,i}$ of the sample is assigned to a partition $j \in [1 \dots k]$.
Accordingly, its Voronoi cell with respect to $\set{P}_S$ defines the sub-space
of $\mathbb{R}^d$ associated with partition $j$.
In order to extend the partitioning to the entire input point set,
each point $p \in \set{P}$
is assigned to the partition of its containing Voronoi cell.

All steps \iftoggle{tr}{in \algref{alg:partition}}{of the partitioning} can be efficiently parallelized.
\citet{sampling} present an efficient parallel random sampling algorithm.
The triangulation of the sample point set $\set{P}_S$ could be computed in
parallel using our DT algorithm recursively.
However, as the sample is small, a fast sequential algorithm is typically more efficient.
Graph conversion is trivially done in parallel and
\citet{parhip} present a parallel graph partitioning algorithm.
The parallelization of the assignment of input points to their respective
partitions
is \iftoggle{tr}{explicitly given in \algref{alg:partition}}{trivial}.%

\iftoggle{tr}{
\begin{table}[tb]
  \centering
  \begin{tabular}{lc}
    \toprule
    \tlhead{Weight} & \thead{$\mathbf{\bm{\omega}(e = (v,w)}$} \\
    \midrule
    constant & 1 \\
    inverse & $\frac{1}{d(v,w)}$ \\
    logarithmic & $-\log d(v,w)$ \\
    linear & $1 - d(v,w)$ \\
    \bottomrule
  \end{tabular} 
  \caption{Possible choices for the edge weight $\omega$,
    with $d(v,w) = \frac{||v - w||}{d^\ast}$ denoting the normalized
Euclidean distance of points $v$ and $w$, with $d^\ast$ being the length of the
maximum diagonal.}
  \label{tab:weights}
\end{table}
} 

\subsection{Recursive Bisection \& Direct \texorpdfstring{$k$}{k}-way Partitioning}\label{sec:strategies}

Two possible strategies exist to obtain $k$ partitions from a graph:
direct $k$-way partitioning and recursive bisection.
For the latter, the graph is recursively partitioned into
$k^\prime = 2$ partitions $\log k$ times.
In the graph partitioning community,
\citet{gpTheory} prove that recursive bisection can lead to arbitrarily bad
partitions and~\citet{gpPractice} confirm the superiority of direct $k$-way
partitioning experimentally.
However, recursive bisection is still widely -- and successfully -- used in
practice
(\eg\iftoggle{tr}{in METIS~\citep{metis} and}{} for initial partitioning in KaHIP~\cite{kahip}).
\iftoggle{tr}{Other problem domains also apply recursive bisection successfully.
In hypergraph partitioning, it can lead to better partitionings in the presence
of large hyperedges, \ie edges with many vertices~\citep{kahypar}.}{}
We therefore consider both partitioning variants for obtaining $k$ partitions
for our DT algorithm.

\begin{figure}[tb]
	\centering
	\begin{subfigure}{0.25\textwidth}
		\centering
		\includegraphics[width=1.0\textwidth,keepaspectratio]{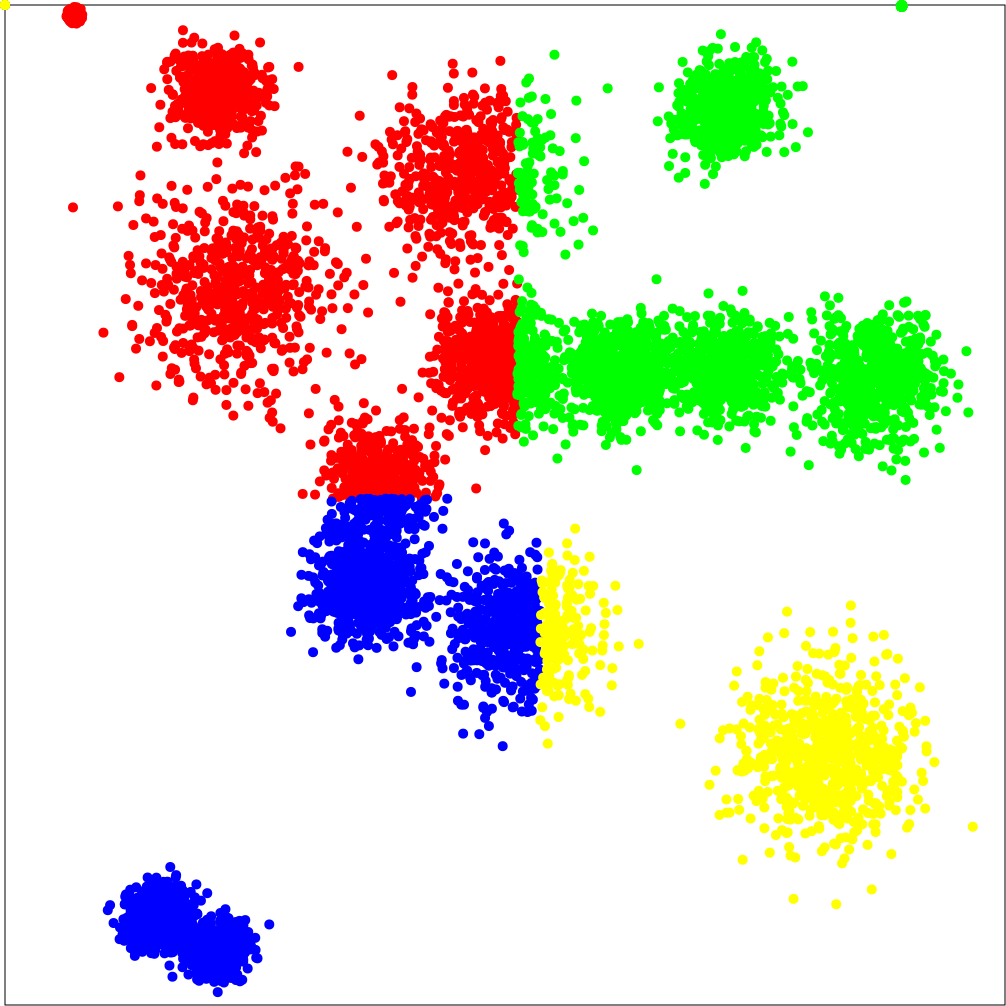}
		\caption{cycle}
		\label{fig:patchbubbles:cycle}
	\end{subfigure}
	\begin{subfigure}{0.25\textwidth}
		\centering
		\includegraphics[width=1.0\textwidth,keepaspectratio]{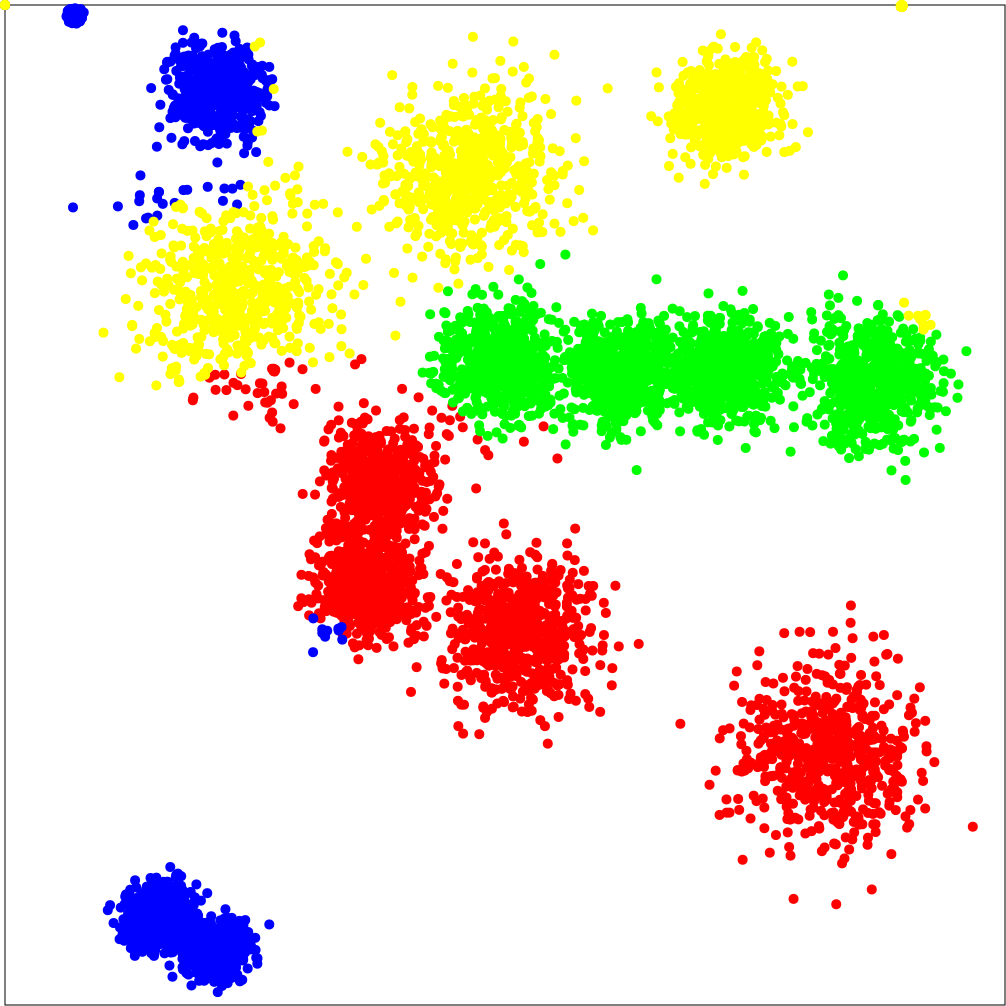}
		\caption{direct $k$-way}
		\label{fig:patchbubbles:kway}
	\end{subfigure}
	\begin{subfigure}{0.25\textwidth}
		\centering
		\includegraphics[width=1.0\textwidth,keepaspectratio]{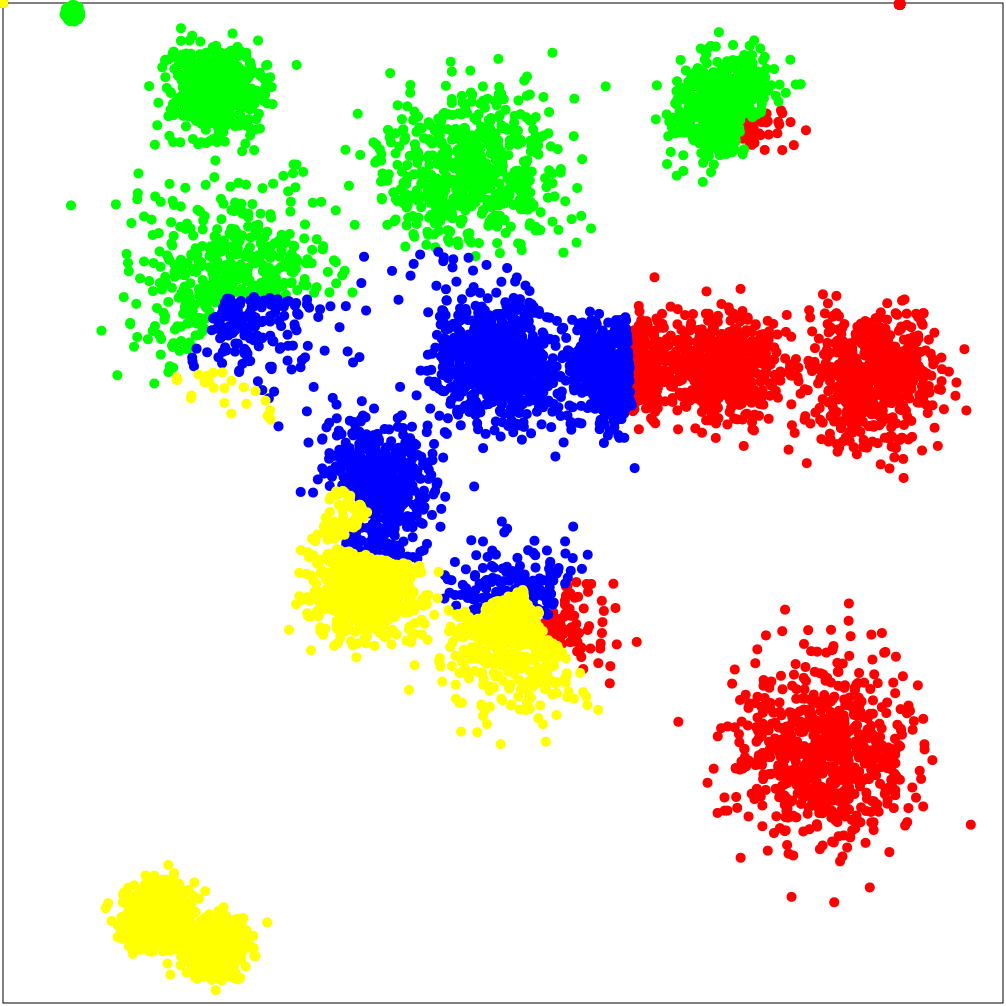}
		\caption{recursive bisection}
		\label{fig:patchbubbles:bi}
	\end{subfigure}
	\caption{Example for two-dimensional partitioning with \num{10000} points and
    a sample size of \num{1000}.}
	\label{fig:patchbubbles}
\end{figure}

The partitioning schemes originally proposed in~\citep{Funke2017} can be
seen as recursive bisection:
the input is recursively split along the median.
The splitting dimension is chosen in a cyclic fashion, similiar to $k$-D trees.
\figref{fig:patchbubbles:cycle} shows an example.

Similarly, \iftoggle{tr}{\algref{alg:partition}}{our new partitioning algorithm} can be applied $\log k$ times,
at each step $i$ drawing a new sample point set $\set{P}_{S, i}$,
triangulating and partitioning $\set{P}_{S,i}$,
and assigning the remaining input points to their respective partition.
As in the original scheme,
this leads to $k-1$ merge steps,
entailing $k-1$ border triangulations.
In the sample-based approach however,
the partitioning avoids cutting dense regions of the input,
which would otherwise lead to large and expensive border triangulations;
refer to \figref{fig:patchbubbles:bi}.

Using direct $k$-way partitioning,
only one partitioning and one merge step is required.
The single border point set will be larger,
with points spread throughout the entire input area.
This however, allows for efficient parallelization of the border triangulation
step using our DT algorithm recursively.
\figref{fig:patchbubbles:kway} depicts an example partitioning.


\subsection{Geometric Primitives}\label{sec:geo}

Our \dac algorithm~\citep{Funke2017} mostly relies on combinatorial
computations on hash values except for the base case computations and the
detection of the border simplices.
The original partitioning schemes always result in partitions
defined by axis-aligned bounding boxes.
Therefore,
\iftoggle{tr}
{the intersection test in \lnref{ln:intersectionTest} in \algref{alg:sma}}
{the test whether the circumhypersphere of a simplex intersects another partition}
can be performed using the fast box-sphere overlap test of
\citet{circletest}.
However, using the more advanced partitioning algorithms presented in this
paper, this is no longer true.
Therefore the geometric primitives to determine the border simplices need to be
adapted to the more complexly shaped partitions.
The primitives need to balance the computational cost of the
intersection test itself with the associated cost for including non-essential
points in the border triangulation.

\iftoggle{tr}{
\begin{figure}[tb]
    \centering
    \begin{subfigure}[t]{0.24\textwidth}
        \centering
        \includegraphics[width=.95\textwidth, page=1]{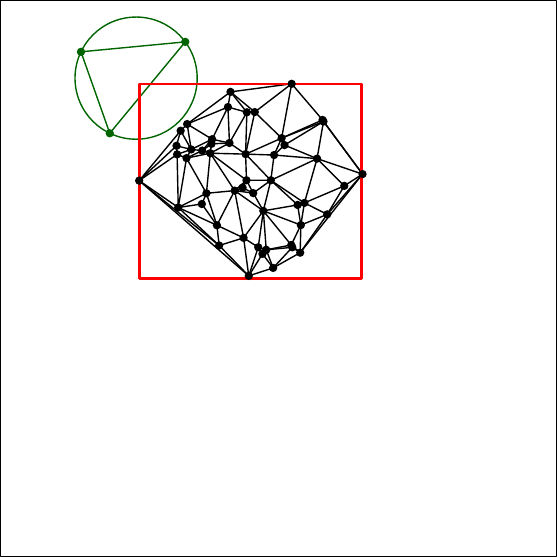}
        \caption{bounding box}
        \label{fig:geo:bbox}
    \end{subfigure}%
    \begin{subfigure}[t]{0.24\textwidth}
        \centering
        \includegraphics[width=.95\textwidth, page=3]{images/geo}
        \caption{grid-based}
        \label{fig:geo:grid}
    \end{subfigure}
    \begin{subfigure}[t]{0.24\textwidth}
        \centering
        \includegraphics[width=.95\textwidth, page=4]{images/geo}
        \caption{exact}
        \label{fig:geo:exact}
    \end{subfigure}
    \caption{Partition boundary determination strategies.
      The path through the AABB tree to test for intersection with the circle in the upper left is
    marked by the colored squares. The tested points for the exact strategy are highlightes in red.}
    \label{fig:geo}
\end{figure}
}

\iftoggle{tr}{%
\subsubsection{Bounding Box Intersection Test}\label{sec:bbox}

A crude approximation uses the bounding box of each partition and the fast
intersection test of~\citet{circletest} to determine the simplices that belong
to the border of a partition.
While computationally cheap, the bounding box can overestimate the extent of a partition.
\iftoggle{tr}{\figref{fig:geo:bbox} provides an example.}

\subsubsection{Grid-based Intersection Test}\label{sec:aabb}

To improve accuracy while still keeping the determination of the border simplices geometrically simple
and computationally cheap,
we use a uniform grid combined with an AABB tree~\citep{aabb}.
For each partition $\set{P}_k$ it is determined which cells of the uniform grid $\mathcal{G}$ are occupied by
points from that partition,
\ie, $\set{\mathcal{C}}_k = \left\{c \in \mathcal{G}: \exists p \in \set{P}_k: p \in c \right\}$.
To accelerate the intersection tests we build an AABB tree on top of each set
$\set{\mathcal{C}}_k$\iftoggle{tr}{, depicted in \figref{fig:geo:grid}}{}.
The AABB tree is built once for every partition $k$ and contains the occupied grid
cells $\set{\mathcal{C}}_k$ as leaves and recursively more coarse-grained
bounding boxes.
The root node of the tree corresponds to the bounding box from \secref{sec:bbox}.
This allows for a more accurate test whether a given simplex $s$ of
partition $i$ intersects with partition $j$ using $\log |\set{\mathcal{C}}_j|$ box-sphere
intersection tests~\citep{circletest}.

\subsubsection{Exact Intersection Test}

In order to only add the absolutely necessary points to the border triangulation
an even more computationally expensive test is required.
For a given simplex $s$ of partition $i$ we use the AABB intersection test from the previous section
to determine the set $\set{\mathcal{C}}^\prime_j \subseteq \set{\mathcal{C}}_j$ of cells intersected
by the circumhypersphere of $s$ in partition $j$.
For all points contained in these cells an adaptive precision \emph{inSphere}-test \citep{Shewchuk1997} is performed to determine
whether $s$ violates the Delaunay property and thus its vertices need to be added to the border triangulation.
}{
We propose three intersection tests:\footnote{For a more detailed description of the primitives we refer to the technical report~\citep{fullpaper}.}
\begin{enumerate*}[i)]
\item each partition is crudely approximated with an axis-aligned \emph{bounding box}
and the fast intersection test of~\citet{circletest} is used to determine the simplices that belong to the border of a partition.
While computationally cheap, the bounding box can overestimate the extent of a partition.
\item for each partition it is determined which cells of a \emph{uniform grid} are occupied by points from that partition.
This allows for a more accurate test whether a given simplex $s$ of
partition $i$ intersects with partition $j$ by determining whether any of $j$'s occupied grid cells are intersected by circumhypersphere of $s$,
again using the box-sphere intersection test~\citep{circletest}.
To further accelerate the intersection test we build an AABB tree~\citep{aabb} on top of the grid data structure.
\item to \emph{exactly} determine the necessary points for the border triangulation we use the previous test to find the grid cells of partition $j$ intersected by the circumhypersphere of $s$ and then use the an adaptive precision \emph{inSphere}-test \citep{Shewchuk1997} for all points contained in these cells to test whether $s$ violates the Delaunay property and thus its vertices need to be added to the border triangulation.
\end{enumerate*}
}

\iftoggle{tr}{
\subsection{Implementation Notes}\label{sec:impl}

We integrated our divide-step into the implementation of~\citep{Funke2017},
which is available as open source.\footnote{%
\url{https://git.scc.kit.edu/dfunke/DelaunayTriangulation}}
We use KaHIP~\citep{kahip} and its parallel version~\citep{parhip} as graph
partitioning tool.
The triangulation of the sample point set is computed sequentially using
CGAL~\citep{cgal:hs-chdt3-15b} with exact predicates.\footnote{%
\texttt{CGAL::Exact\_predicates\_inexact\_constructions\_kernel}}
The closest sample point for a given input point in \lnref{ln:partition:nn}
of \algref{alg:partition} can be found via the Voronoi diagram of the sample
triangulation.
However, using the lightweight $k$-D tree implementation
\emph{nanoflann}\footnote{%
\url{https://github.com/jlblancoc/nanoflann}} proved to be more efficient.
}

\section{Evaluation}\label{sec:eval}

\iftoggle{tr}{
\begin{table}[tb]
  \centering
  \begin{tabular}{lrrrr}
    \toprule
    \tlhead{Distribution} & \thead{Points} & \thead{Simplices} & \thead{$\frac{\text{simplices}}{\text{point}}$} & \thead{Runtime} \\
    \midrule
    uniform     & \num{50000000}   & \num{386662755} & \num{7.73}    & \SI{164.6} {\second} \\
    normal      & \num{50000000}   & \num{390705843} & \num{7.81}    & \SI{162.6} {\second} \\
    ellipsoid   & \num{500000}     & \num{23725276}  & \num{4.74}    & \SI{88.6}{\second} \\
    lines       & \num{10000}      & \num{71540362}  & \num{7154.04} & \SI{213.3}{\second} \\
    bubbles     & \num{50000000}   & \num{340201778} & \num{6.80}    & \SI{65.9}{\second} \\
    malicious   & \num{50000000}   & \num{143053859} & \num{2.86}    & \SI{63.9}{\second} \\
    \cmidrule(lr){1-5}
    Gaia DR2    & \num{50000000}   & \num{359151427} & \num{7.18} & \SI{206.9}{\second} \\
    \bottomrule
  \end{tabular}
  \caption{Input point sets and their resulting triangulations. 
  Running times are reported for $k = t= 16$, parallel KaHIP, $\eta(n) = \sqrt{n}$ and logarithmic edge weights.}
  \label{tab:pointsets}
\end{table}
}

\iftoggle{tr}{
\begin{figure}[tb]
  \centering
  \includegraphics[width=.5\textwidth]{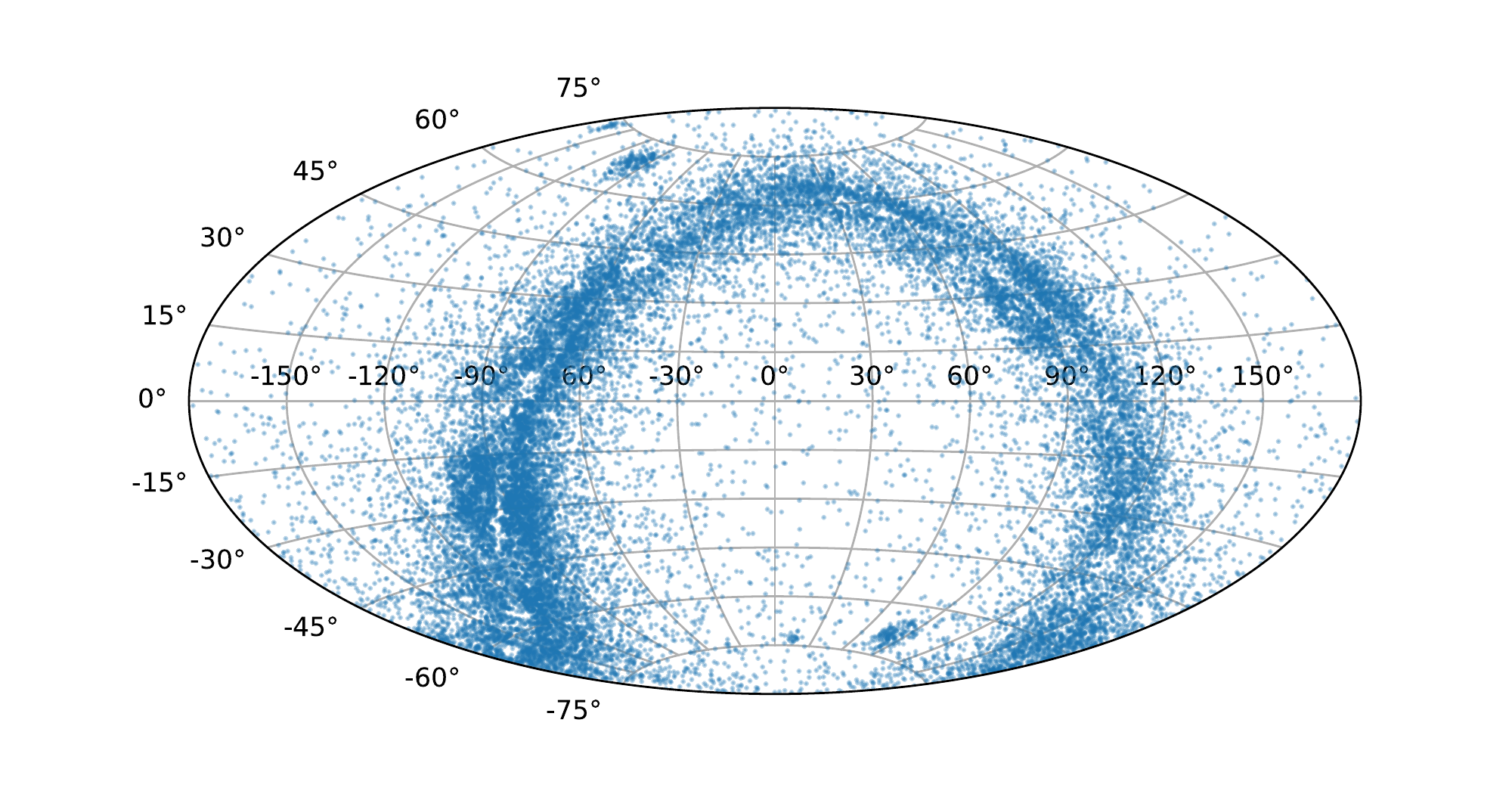}
  \caption{Aitoff projection of a random sample of \num{25000} sources from the
  Gaia DR2 dataset.}
  \label{fig:gaia}
\end{figure}
}

\citet{Batista2010} propose three input point distributions to evaluate 
the performance of their DT algorithm:
$n$ points distributed uniformly
\begin{enumerate*}[a)]
  \item in the unit cube;
  \item on the surface of an ellipsoid; and
  \item on skewed lines.
\end{enumerate*}
Furthermore,~\citet{Lee2001} suggest normally distributed input points around
\begin{enumerate*}[a),resume]
  \item the center of the unit cube; and
  \item \label{dist:bubb} several points within the unit cube -- called \enquote{bubbles}.
\end{enumerate*}
We study two variants of distribution \ref{dist:bubb} with the bubble centers:
\begin{enumerate*}[i)]
  \item distributed uniformly at random in the unit cube;
  \item along the axes of the cycle partitioner cuts -- called
    \enquote{malicious} distribution.
\end{enumerate*}  
We furthermore test our algorithm with a real world dataset from astronomy.
The Gaia DR2 catalog~\citep{gaiadr2} contains celestial positions and the apparent
brightness for approximately \num{1.7} billion stars.
Additionally, for \num{1.3} billion of those stars, parallaxes and proper
motions are available, enabling the computation of three-dimensional
coordinates.
\iftoggle{tr}%
{As \figref{fig:gaia} shows, t}{T}he data exhibits clear structure,
which can be exploited by our partitioning strategy.
We use a random sample of the stars to evaluate our algorithm.
All experiments are performed in three-dimensional space ($D=3$)
\iftoggle{tr}{}{ with up to \num{50e6} input points}.

\iftoggle{tr}%
{\tabref{tab:pointsets} gives}%
{In the technical report~\citep{fullpaper} we give}
an overview of all input point sets,
along with the size of their resulting triangulation%
\iftoggle{tr}{}{ and a visualization of the Gaia dataset}.

The algorithm was evaluated on a machine with dual
Intel Xeon E5-2683 16-core processors and 512\,GiB of main memory.
The machine is running Ubuntu 18.04, with
GCC version 7.2 and CGAL version 4.11.

\iftoggle{tr}{}{
\emph{Implementation Notes:}
We integrated our divide-step into the implementation of~\citep{Funke2017},
which is available as open source.\footnote{%
\url{https://git.scc.kit.edu/dfunke/DelaunayTriangulation}}
We use KaHIP~\citep{kahip} and its parallel version~\citep{parhip} as graph
partitioning tool.
The triangulation of the sample point set is computed sequentially using
CGAL~\citep{cgal:hs-chdt3-15b} with exact predicates.\footnote{%
\texttt{CGAL::Exact\_predicates\_inexact\_constructions\_kernel}}
}

\subsection{Parameter Studies}\label{sec:eval:params}

\iftoggle{tr}{
\newcounter{savefootnote}
\setcounter{savefootnote}{\value{footnote}}
\setcounter{footnote}{1}
\renewcommand{\thefootnote}{\fnsymbol{footnote}} 
\begin{table}[tb]
    \centering
    \begin{tabular}{ll}
    \toprule
    \tlhead{Parameter} & \tlhead{Values} \\
    \midrule
    sample size $\eta(n)$ & \SI{1}{\percent}, \SI{2}{\percent}, $\log n$, $\sqrt{n}$ \\
    KaHIP configuration & \textsc{strong}, \textsc{eco}, \textsc{fast}, \textsc{parallel} \\
    edge weight $\omega(e)$ & constant, inverse, log, linear\tablefootnote{see \tabref{tab:weights}} \\
    geometric primitive & bbox, exact, grid with cell sizes $c_\mathcal{G} = [\frac{1}{2}, \num{1}, \num{2}]$ \\
    \cmidrule(lr){1-2}
    partitions $k$ & $1,2,4,\dots,64$ \\
    threads $t$ & $t = k$ \\
    points $n$ & $[\num{1}, \num{5}, \num{10},  \num{25}, \num{50}] \cdot \num{e6}$\tablefootnote{unless otherwise stated in \tabref{tab:pointsets}} \\
    distribution & see \tabref{tab:pointsets} \\
    \bottomrule
    \end{tabular}
    \caption{Parameters of our algorithm (top) and the conducted experiments (bottom).}
    \label{tab:params}
\end{table}
\setcounter{footnote}{\value{savefootnote}}
\renewcommand{\thefootnote}{\arabic{footnote}}
}

\iftoggle{tr}{
The parameters listed in \tabref{tab:params} can be distinguished into configuration parameters of our algorithm
and parameter choices for our experiments.
In the following we examine the configuration parameters and determine robust choices for all inputs.
The parameter choice influences the quality of the partitioning with respect to partition size deviation and
number of points in the border triangulation.
As inferior partitioning quality will result in higher execution times, 
we use it as indicator for our parameter tuning.
Even though choices for the parameters are correlated,
we present each parameter individually for clarity.
We use the uniform, normal, ellipsoid and random bubble distribution for our parameter tuning
and compare against the originally proposed cyclic partitioning scheme for reference.

\subsection{Sample Size}\label{sec:eval:ss}

\begin{figure}[tbp]
	\centering
	\includegraphics[width=.8\textwidth,keepaspectratio]{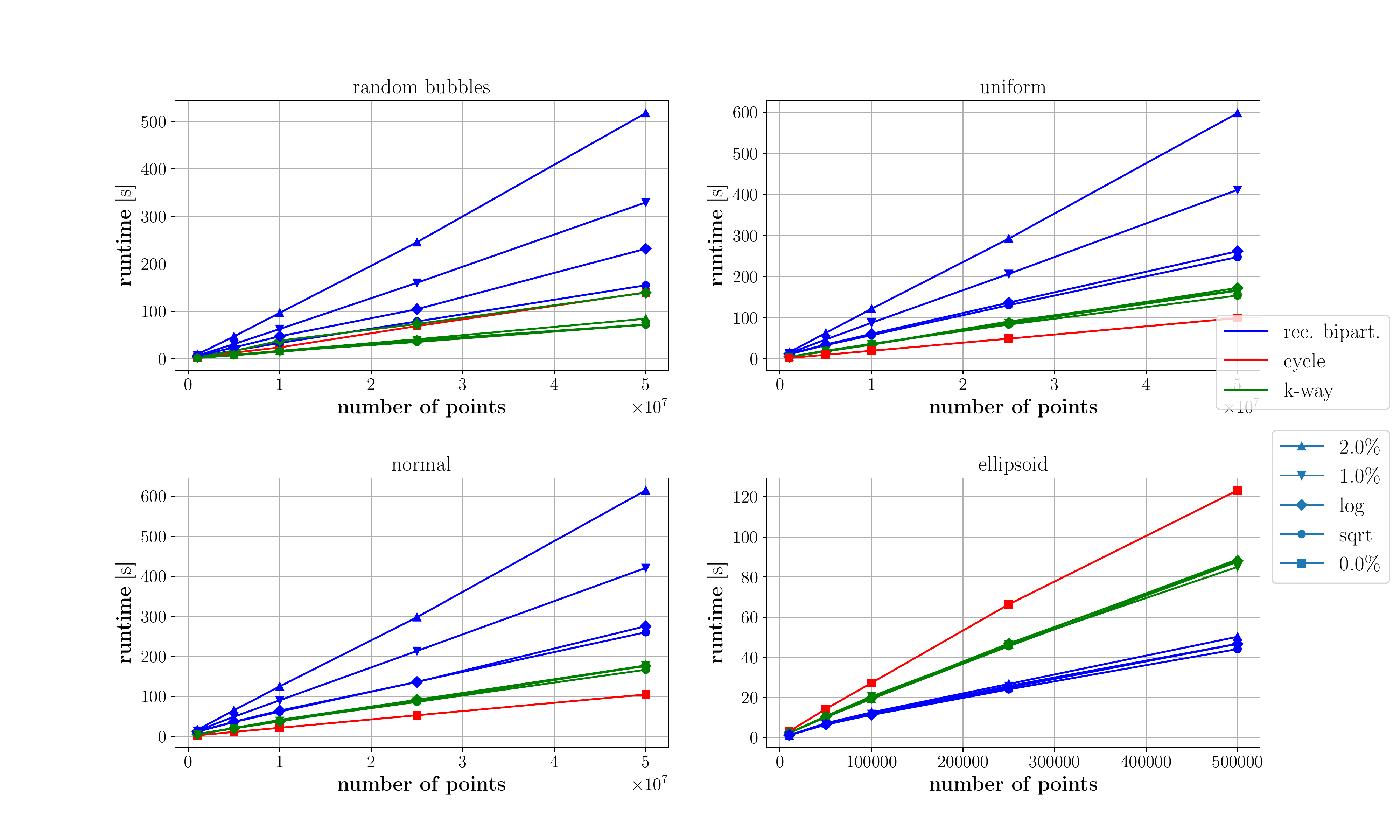}
	\caption{Sample size experiments with $k = t= 16$, logarithmic edge weights, grid-based intersection test with $c_\mathcal{G} = 1$ and parallel KaHIP.}
	\label{fig:ss}
\end{figure}

The main goal of our divide-step is to approximate a good partitioning of the final triangulation of $DT(\set{P})$.
Clearly, a larger sample size $\eta(n)$ yields a better approximation at the cost of an increased runtime for the sample triangulation.
On the other hand, a higher partitioning quality results in better load-balancing between partitions and smaller border triangulations.
\figref{fig:ss} shows the total triangulation time for various choices of $\eta(n)$ for a fixed choice of edge weight and KaHIP configuration.
The runtime of our $k$-way strategy shows little dependence on the sample size,
whereas for recursive bisection the higher runtime for larger sample triangulations clearly outweighs any benefit gained from a better partitioning.
We therefore choose $\eta(n) = \sqrt{n}$ as default for all subsequent experiments.

\subsection{Partitioner Configuration}\label{sec:eval:part}

\begin{figure}[tbp]
	\centering
	\includegraphics[width=.8\textwidth,keepaspectratio]{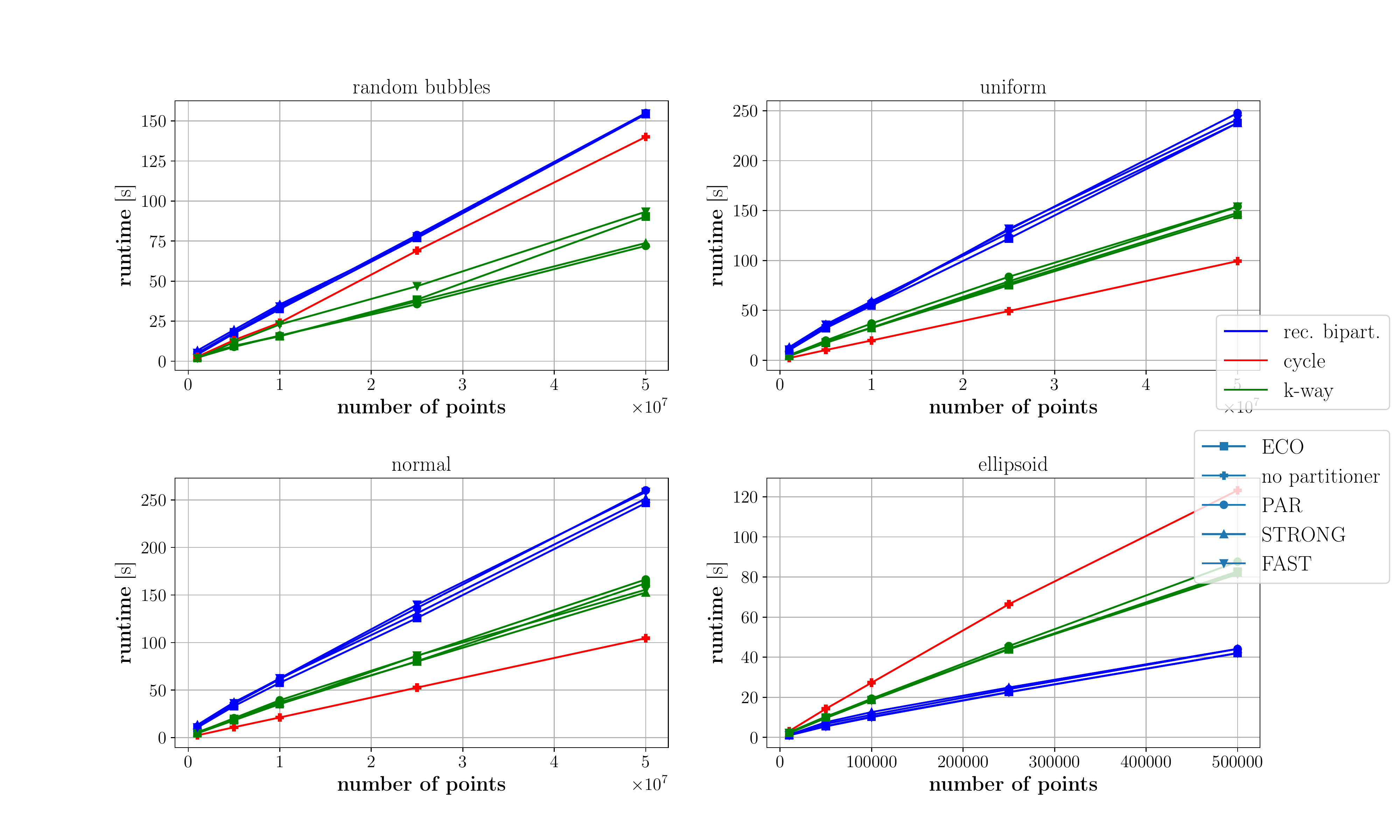}
	\caption{KaHIP configuration experiments with $k = t= 16$, logarithmic edge weights, grid-based intersection test with $c_\mathcal{G} = 1$ and $\eta(n) = \sqrt{n}$.}
	\label{fig:part}
\end{figure}


Numerous configuration parameters balance quality and runtime in graph partitioning~\citep{kahip}.
KaHIP defines several presets of its parameters, each providing a good trade-off for a given runtime or quality requirement;
these are, with increasing focus on runtime: \textsc{strong}, \textsc{eco} and \textsc{fast}~\citep{kahip_params}.
Additionally, a set of parameters specifically tuned for social and web graphs is provided.
The shared memory parallel version of KaHIP builds upon these configuration presets and
extends them with parallel algorithms.
The configuration identified as \textsc{parallel} in our experiments corresponds to \textsc{fastsocialmultitry\_parallel} in~\citep{parhip}.
In all experiments, we set the imbalance parameter for KaHIP to $\epsilon = \SI{5}{\percent}$.
\figref{fig:part} shows the total triangulation time for the various KaHIP presets for a fixed choice of edge weight and sample size.
In general, the time taken by the graph partitioning algorithm is very small compared to the DT computations.
Therefore, we expect the runtime to be a direct reflection of the graph partitioning quality.
Our experiments confirm this notion.
For instance, for the random bubble distribution the inferior partition quality of the faster \textsc{eco} preset compared to \textsc{strong} leads 
to an increase of triangulated points of \SI{3.6}{\percent} at the gain of \SI{1.5}{\second} in runtime
-- \SI{2.4}{\percent} of the total runtime.
The parallel KaHIP configuration achieves a similar runtime as \textsc{eco} and only a slightly worse cut than \textsc{strong} (\SI{0.5}{\percent})
and will be the default for all subsequent experiments.

\subsection{Edge Weights}\label{sec:eval:weights}

\begin{figure}[tbp]
	\centering
	\includegraphics[width=.8\textwidth,keepaspectratio]{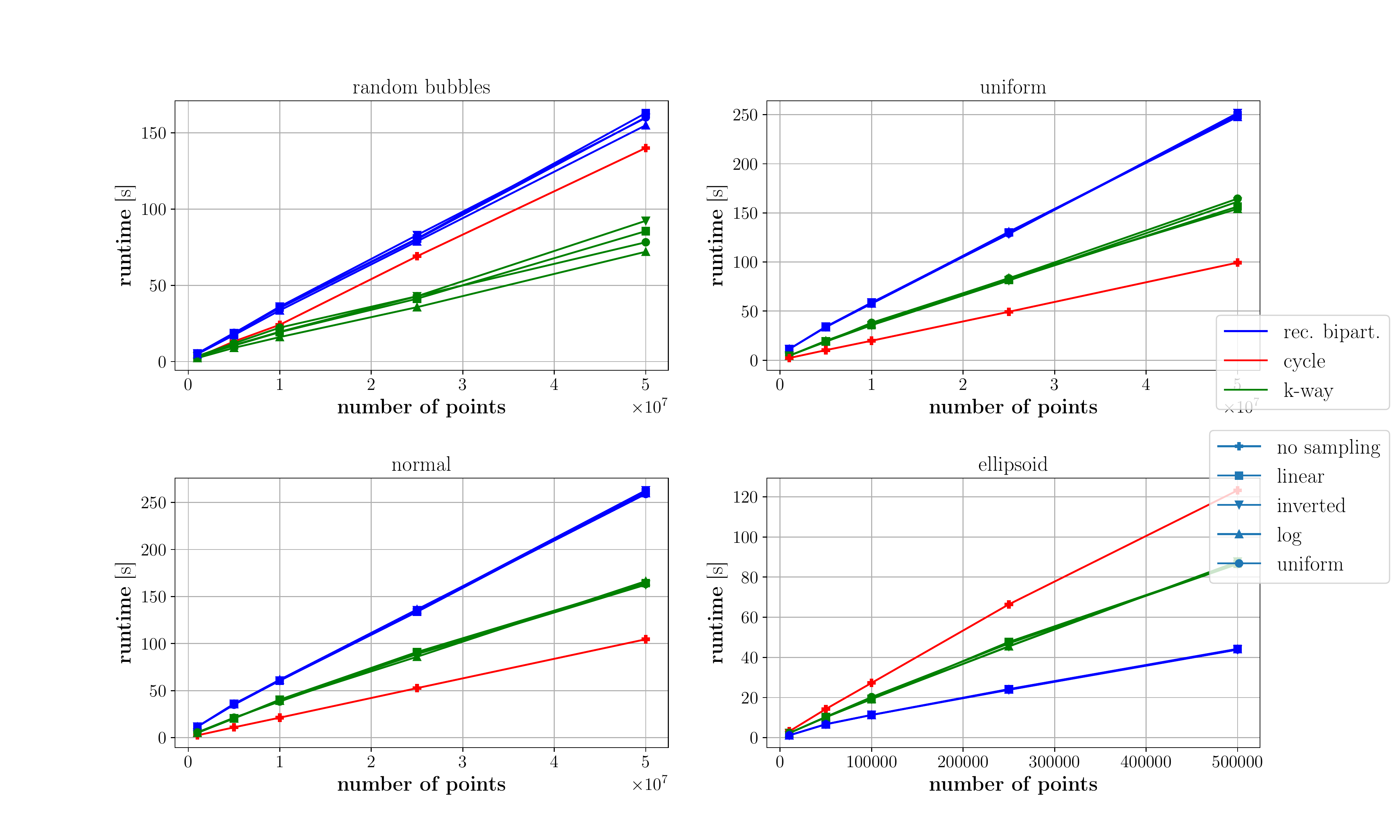}
	\caption{Edge weights experiments with $k = t= 16$, parallel KaHIP, grid-based intersection test with $c_\mathcal{G} = 1$ and $\eta(n) = \sqrt{n}$}
	\label{fig:weights}
\end{figure}


As discussed in \secref{sec:sbp}, 
sparse regions of the input points
-- which are desirable as partition borders --
 result in long Delaunay edges in the sample triangulation.
Since graph partitioning minimizes the weight of the cut edges,
the edge weight needs to be inversely proportional the Euclidean length of the edge,
refer to \tabref{tab:weights}.
\figref{fig:weights} shows the total triangulation time for the various proposed edge weights for a fixed choice of KaHIP configuration and sample size.
As dense regions of the input point set are reflected by \emph{many} short edges in the sample triangulation,
even constant edge weights result in a sensible partitioning.
However, for input distributions with an exploitable structure, such as random bubbles,
logarithmic edge weights lead to \SI{2.3}{\percent} fewer triangulated points,
due to the increased incentive to cut through long -- ergo cheap -- Delaunay edges.

\subsection{Geometric Primitive}\label{sec:eval:is}

\begin{figure}[tbp]
	\centering
	\includegraphics[width=.8\textwidth,keepaspectratio]{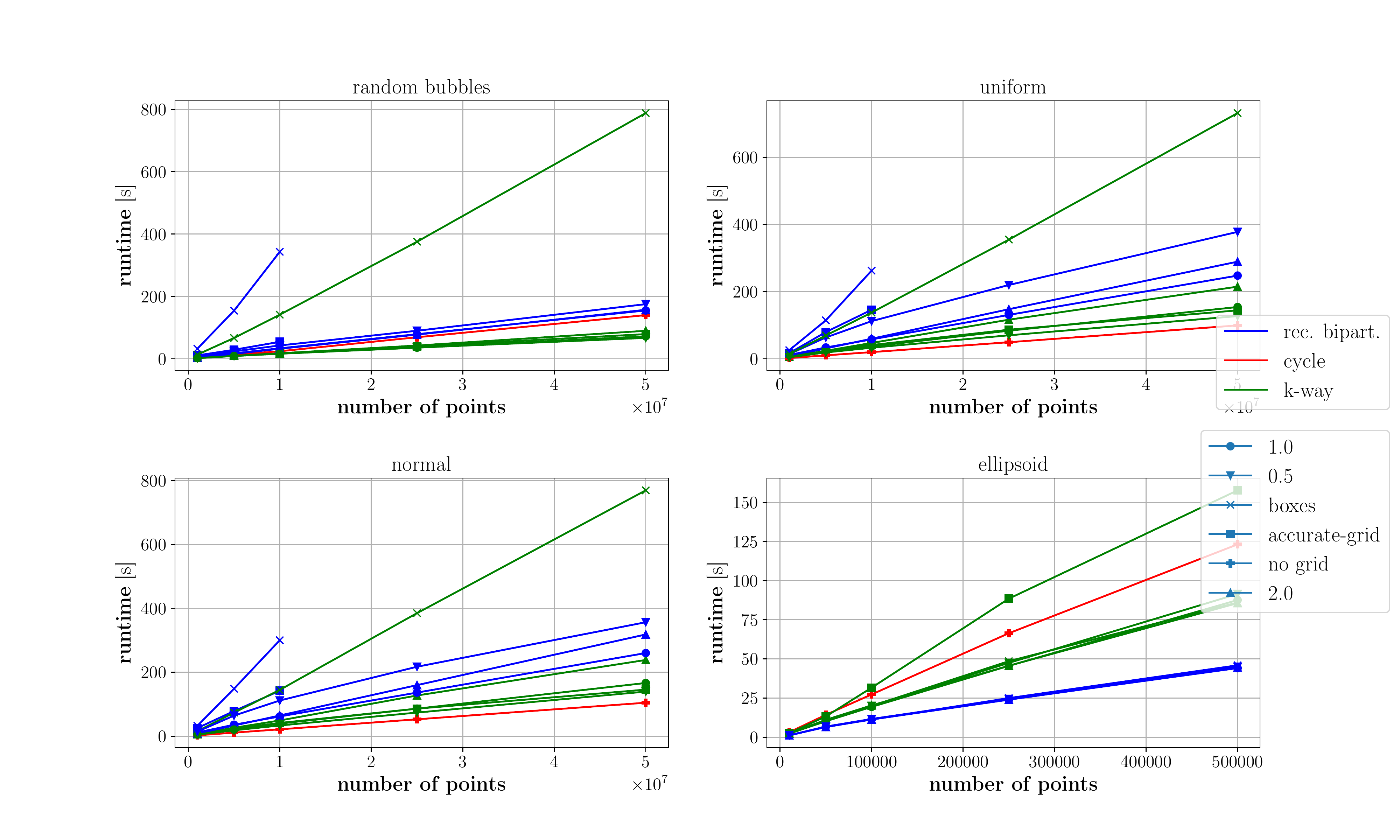}
	\caption{Intersection test experiments with $k = t= 16$, logarithmic edge weights, parallel KaHIP and $\eta(n) = \sqrt{n}$.}
	\label{fig:is}
\end{figure}

The geometric primitive used to determine the border simplices influences both the number of simplices in the border (accuracy)
and the runtime required for the primitive itself.
The intersection tests introduced in \secref{sec:geo} each provide their own trade-off between accuracy and runtime.
The grid-based intersection test requires the grid cell size as further configuration parameter,
which introduces a trade-off between runtime -- mainly memory allocation for the grid data structure -- and accuracy.
\figref{fig:is} shows the total triangulation time for the bounding box, exact and grid-based intersection test,
the latter for various choices of cell size $c_\mathcal{G}$.
The bounding box test produces very large border triangulation and suffers from the resulting runtime penalty.
On the contrary, the exact test produces the smallest border triangulation, the test itself, however, is rather expensive.
The grid-based test provides a good trade-off between the two strategies.
The finer grid better approximates the exact test.
For the uniform and normal distribution the $k$-way strategy clearly profits from the smaller border triangulation,
whereas the effects for distributions with a underlying structure are less pronounced.
The impact of the finer grid on the runtime becomes apparent for the recursive bisection strategy,
which needs to allocate memory repeatedly.
We use the grid-based intersection test with $c_\mathcal{G} = 1$ as default for all subsequent experiments.
} 
{
In our parameter study we examine the configuration parameters of our algorithm and determine robust choices for all inputs.
The parameter choice influences the quality of the partitioning with respect to partition size deviation and
number of points in the border triangulation.
As inferior partitioning quality will result in higher execution times, 
we use it as indicator for our parameter tuning.
We use the uniform, normal, ellipsoid and random bubble distribution for our parameter tuning
and compare against the originally proposed cyclic partitioning scheme for reference.
Due to space constrains we refer to the technical report~\citep{fullpaper} for an in-depth discussion of each parameter individually and
only present a short summary here.

Our experiments show, that a sample size of $\eta(n) = \sqrt{n}$ balances the approximation quality of a partitioning of the final triangulation
with the runtime for the sample triangulation.
Considering edge weights, dense regions of the input point set are reflected by \emph{many} short edges in the sample triangulation.
Therefore, even constant edge weights result in a sensible partitioning.
However, logarithmic edge weights\footnote{%
$\omega(e = (v,w)) = -\log d(v,w)$ width $d(\cdot)$ denoting the Euclidean distance.}
excel in the presence of an exploitable structure in the input points.
For KaHIP we chose the parallel configuration as default as it requires a similar runtime to the \textsc{eco} configuration while achieving a cut only slightly worse then \textsc{strong}.
The grid-based intersection test with a cell size of $c_\mathcal{G} = 1$ shows the best trade-off between accuracy 
-- \ie only essential simplices are included in the border triangulation -- and runtime for the geometric primitive itself.
} 
\subsection{Partitioning Quality}\label{sec:eval:quant}

\iftoggle{tr}{
\begin{figure}[tb]
	\centering
	\begin{subfigure}{\textwidth}
	\centering
        \includegraphics[width=.8\textwidth,keepaspectratio]{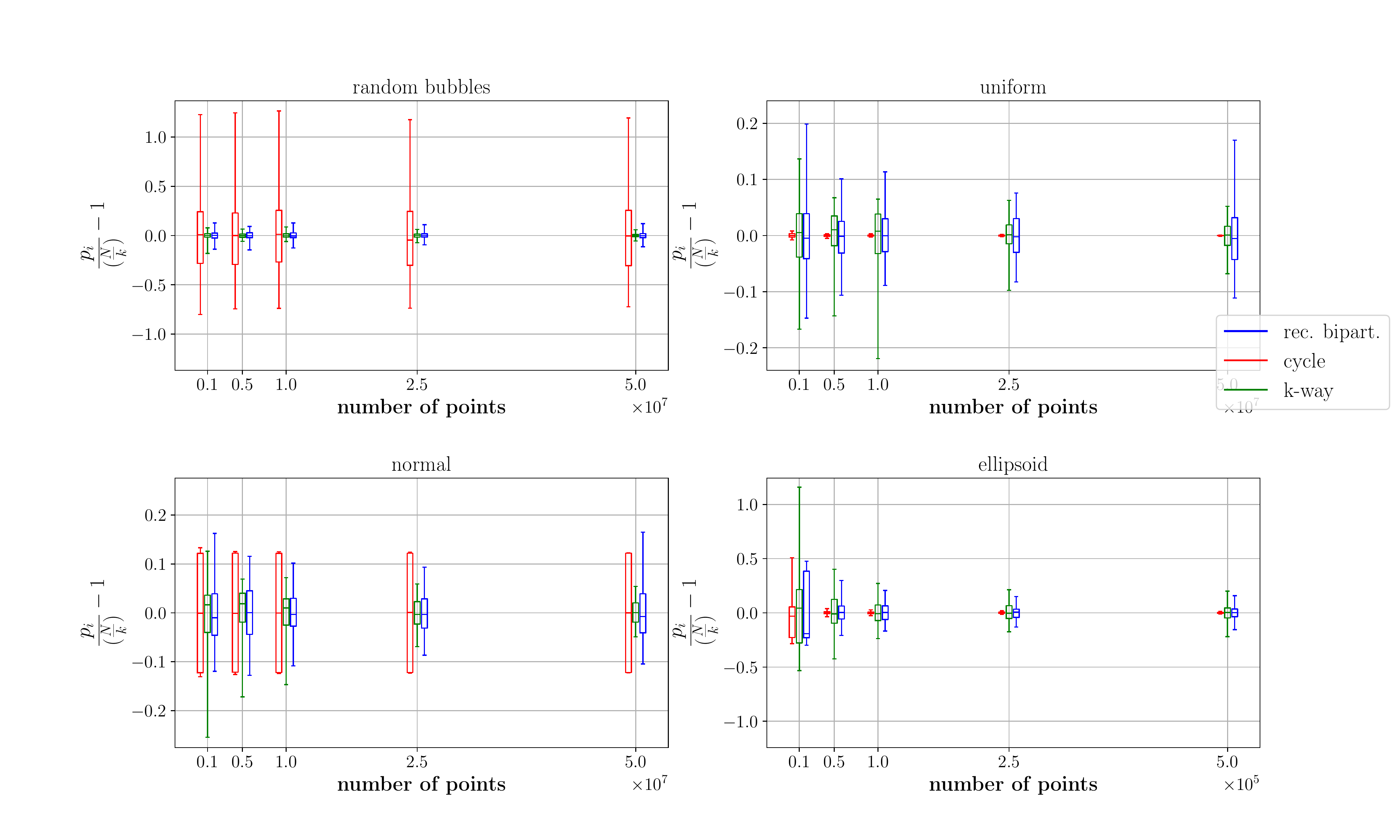}
    \caption{sample size $\eta(n) = \num{.01}n$}
    \label{fig:idev:01}
    \end{subfigure}
    \begin{subfigure}{\textwidth}
    \centering
        \includegraphics[width=.8\textwidth,keepaspectratio]{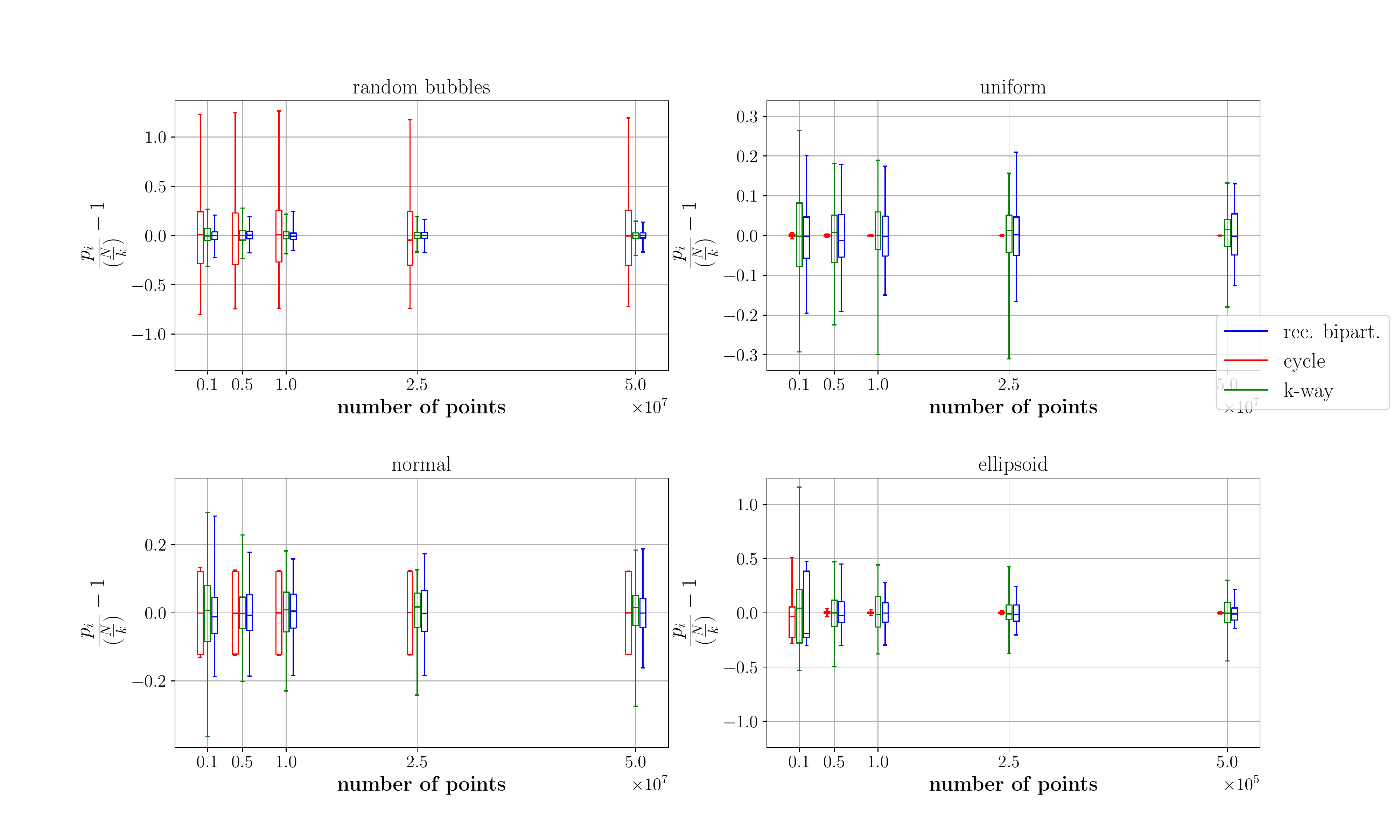}
    \caption{sample size $\eta(n) = \sqrt{n}$}
    \label{fig:idev:sqrt}
    \end{subfigure}
	\caption{Deviation from the ideal partition size for $k = t= 16$, parallel KaHIP, logarithmic edge weights and grid-based intersection test with $c_\mathcal{G} = 1$.}
	\label{fig:idev}
\end{figure}
}

\begin{figure}[tb]
	\centering
	\begin{subfigure}{\textwidth}
	\centering
	\iftoggle{tr}{
        \includegraphics[width=.8\textwidth,keepaspectratio]{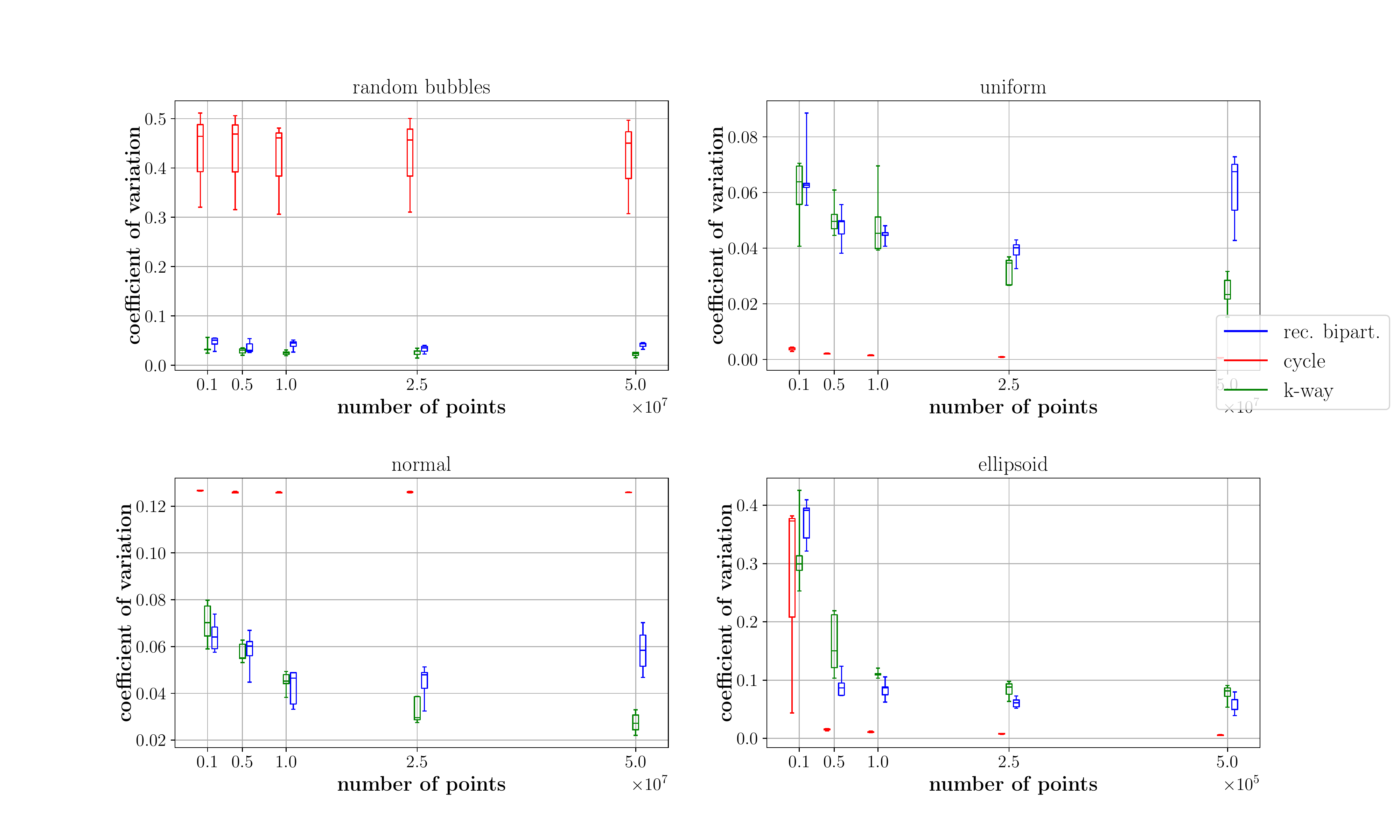}
    }{
        \begin{subfigure}{.49\textwidth}
        \centering
        \includegraphics[width=\textwidth,keepaspectratio]{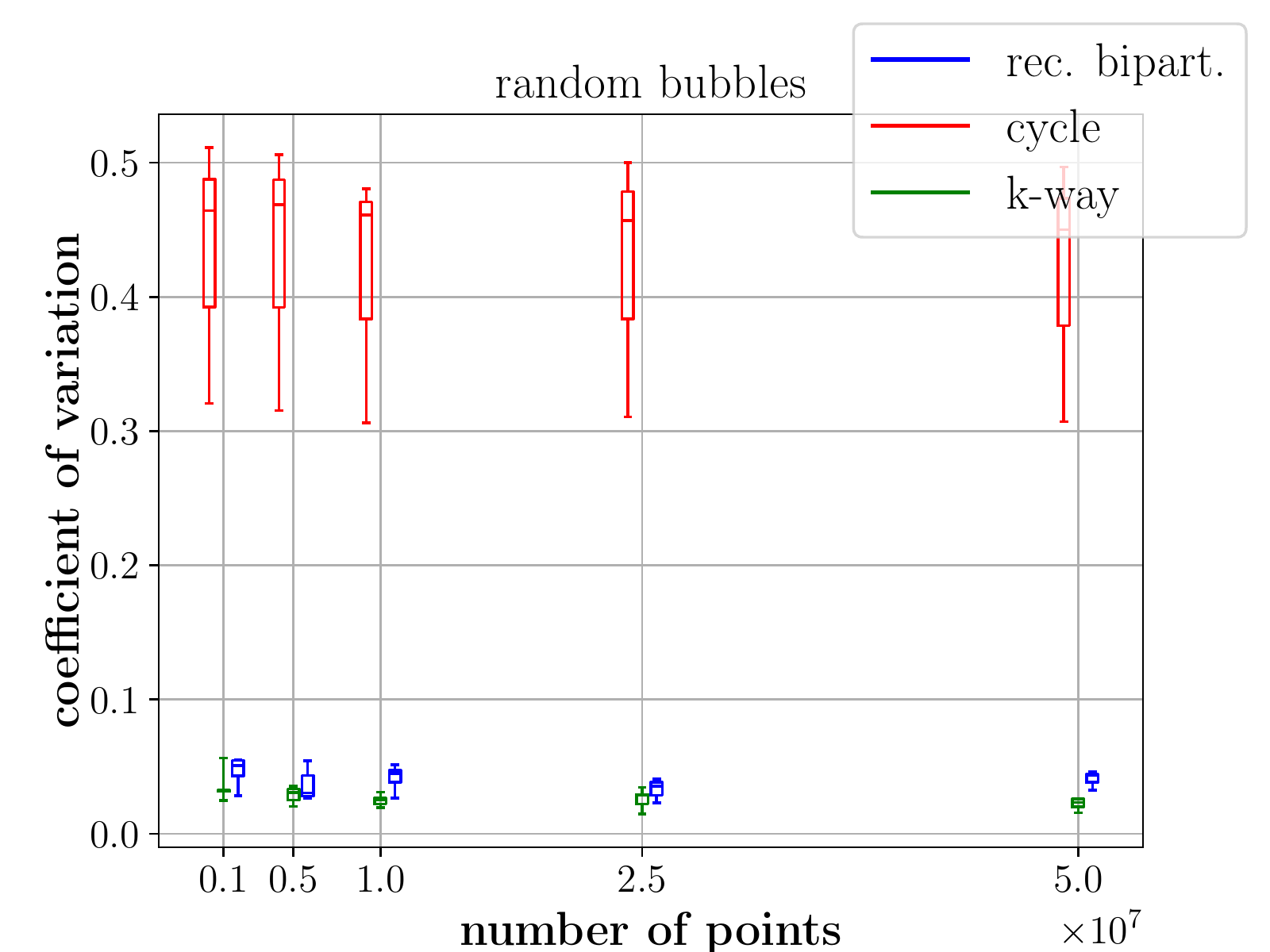}
        \end{subfigure}
        \begin{subfigure}{.49\textwidth}
        \centering
        \includegraphics[width=\textwidth,keepaspectratio]{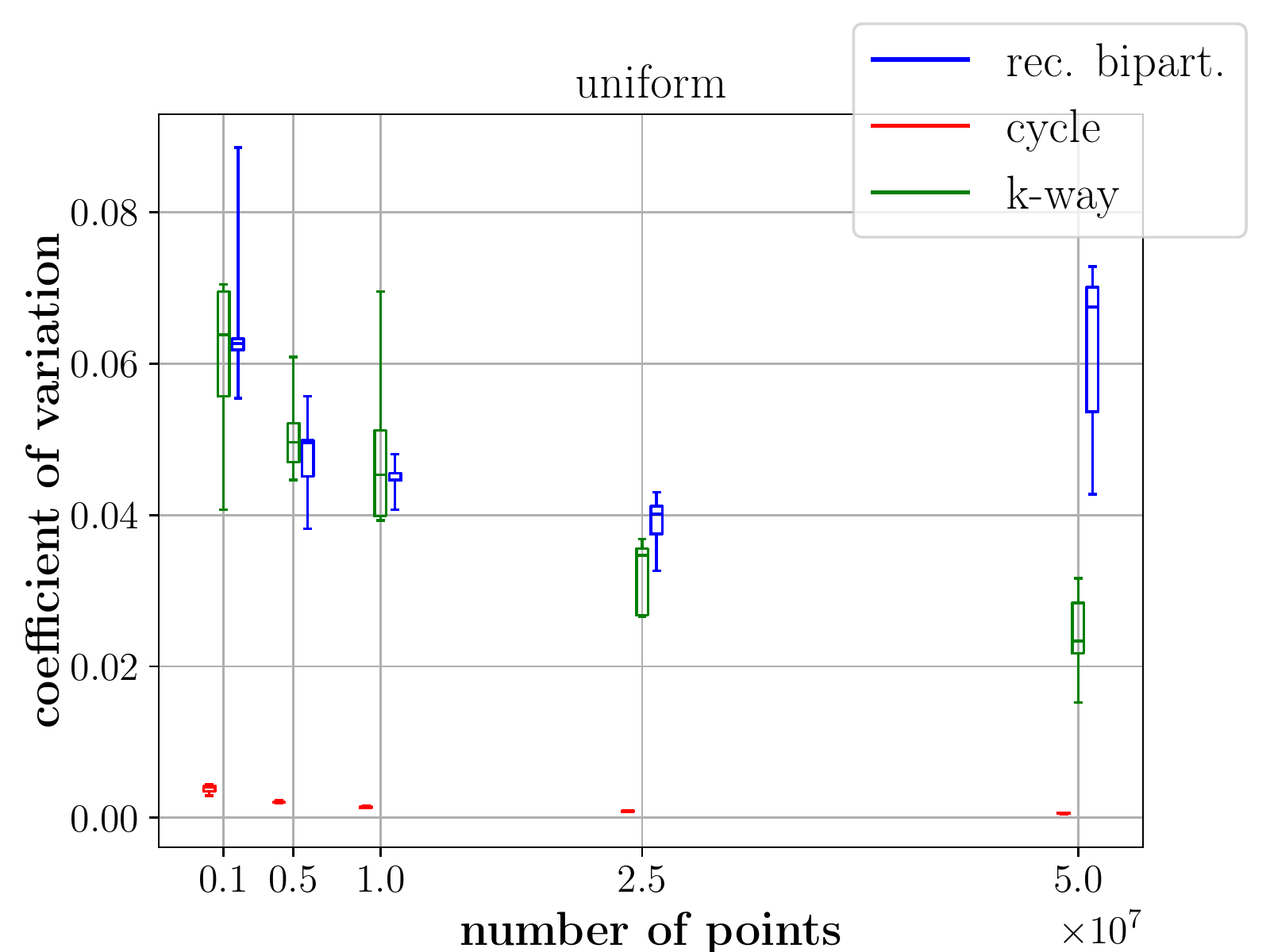}
        \end{subfigure}
    }
    \caption{sample size $\eta(n) = \num{.01}n$}
    \label{fig:dev:01}
    \end{subfigure}
    \begin{subfigure}{\textwidth}
    \centering
    \iftoggle{tr}{
        \includegraphics[width=.8\textwidth,keepaspectratio]{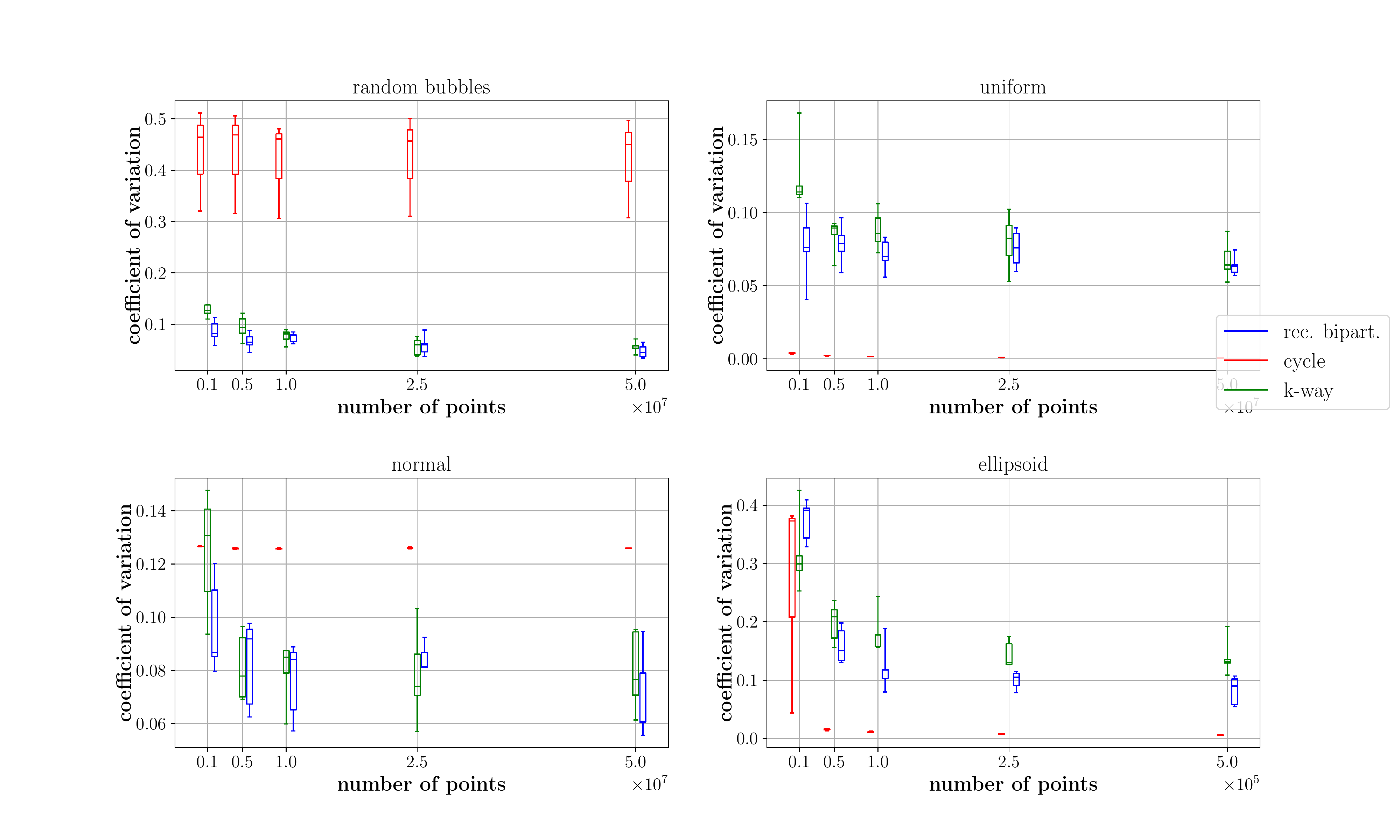}
    }{
        \begin{subfigure}{.49\textwidth}
        \centering
        \includegraphics[width=\textwidth,keepaspectratio]{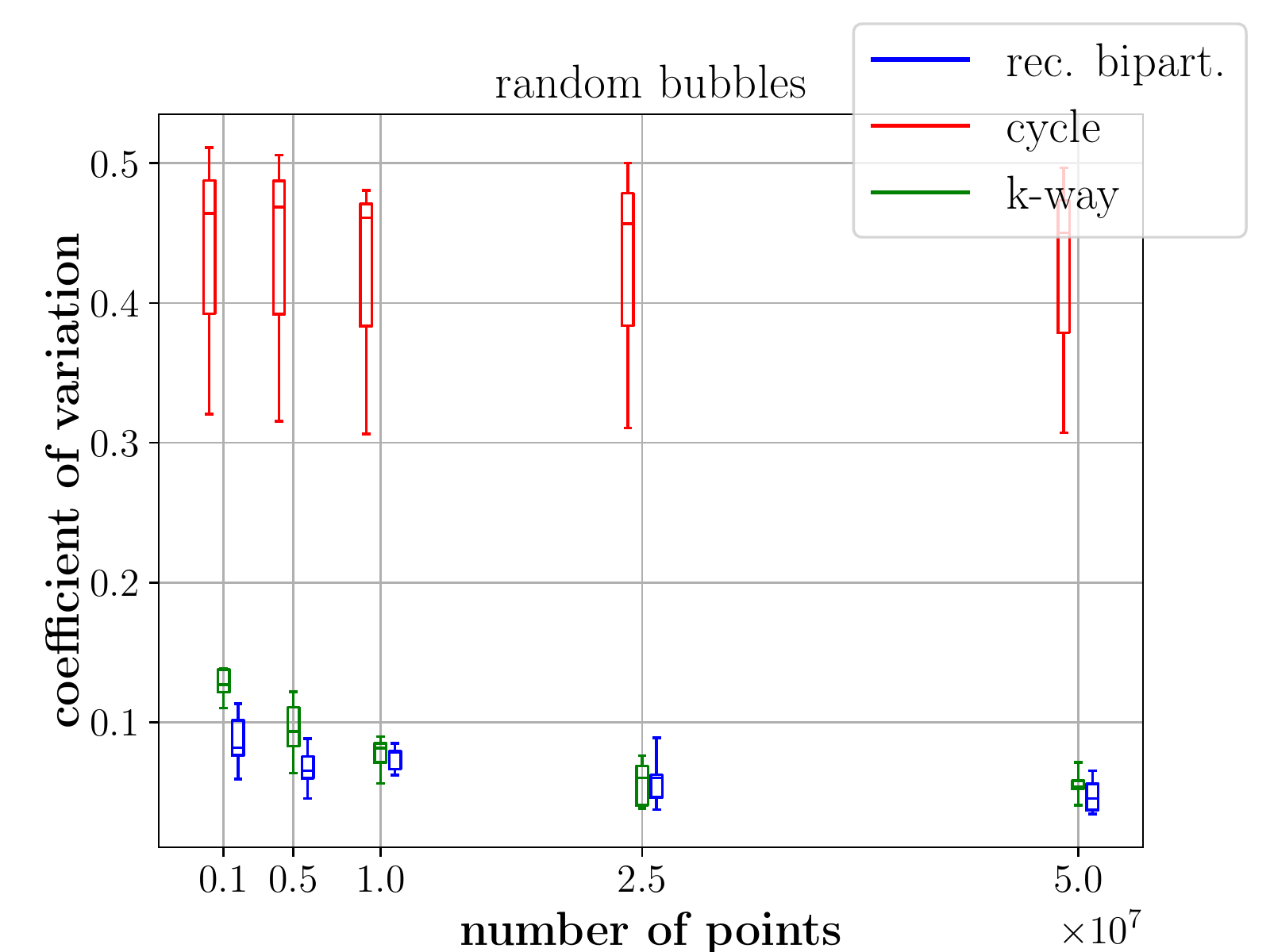}
        \end{subfigure}
        \begin{subfigure}{.49\textwidth}
        \centering
        \includegraphics[width=\textwidth,keepaspectratio]{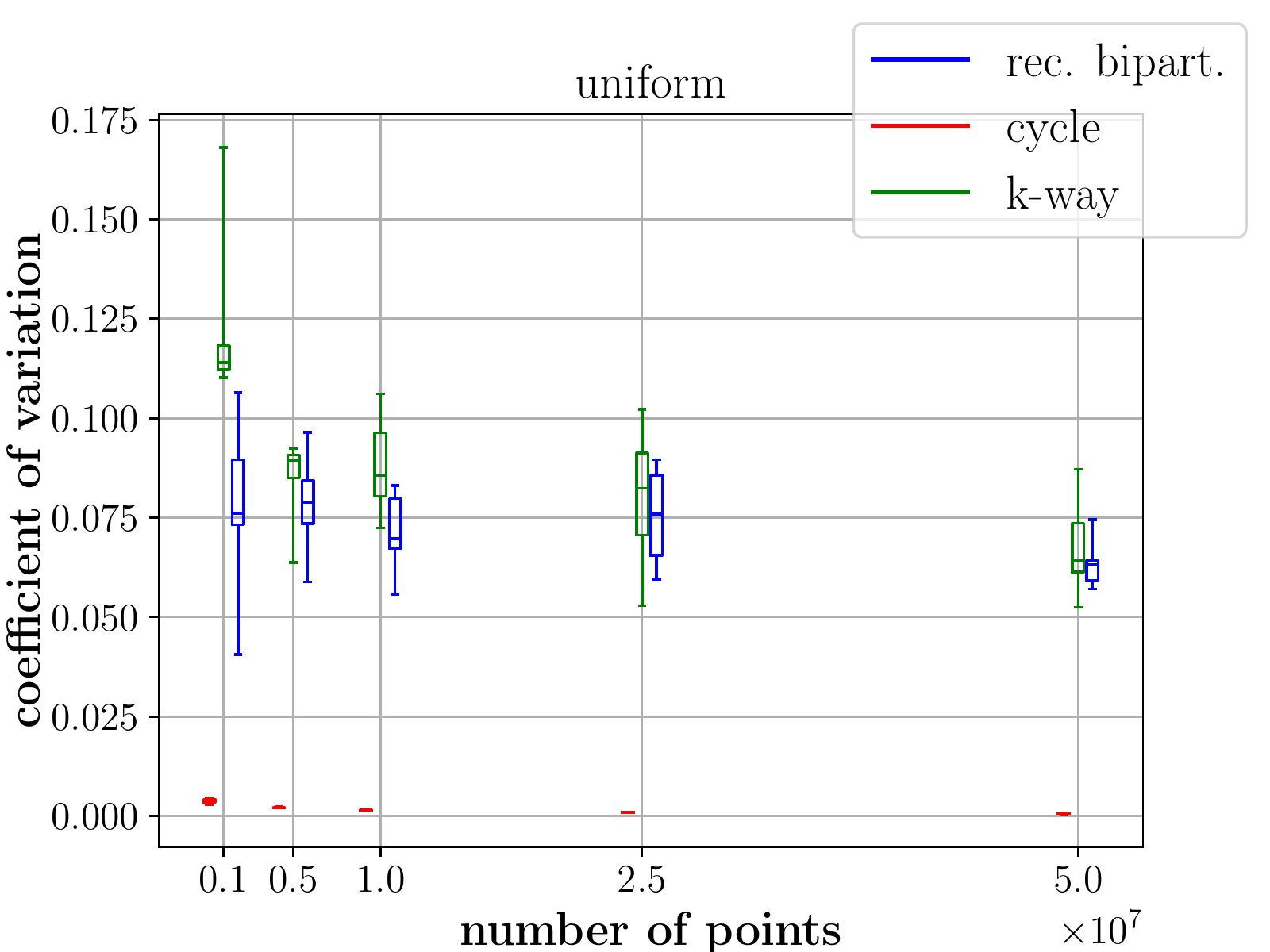}
        \end{subfigure}
    }
    \caption{sample size $\eta(n) = \sqrt{n}$}
    \label{fig:dev:sqrt}
    \end{subfigure}
	\caption{Coefficient of variation of the partition sizes for $k = t= 16$, parallel KaHIP, logarithmic edge weights and grid-based intersection test with $c_\mathcal{G} = 1$.}
	\label{fig:dev}
\end{figure}

\begin{figure}[tb]
	\centering
    \begin{subfigure}{\textwidth}
    \centering
    \iftoggle{tr}{
        \includegraphics[width=.8\textwidth,keepaspectratio]{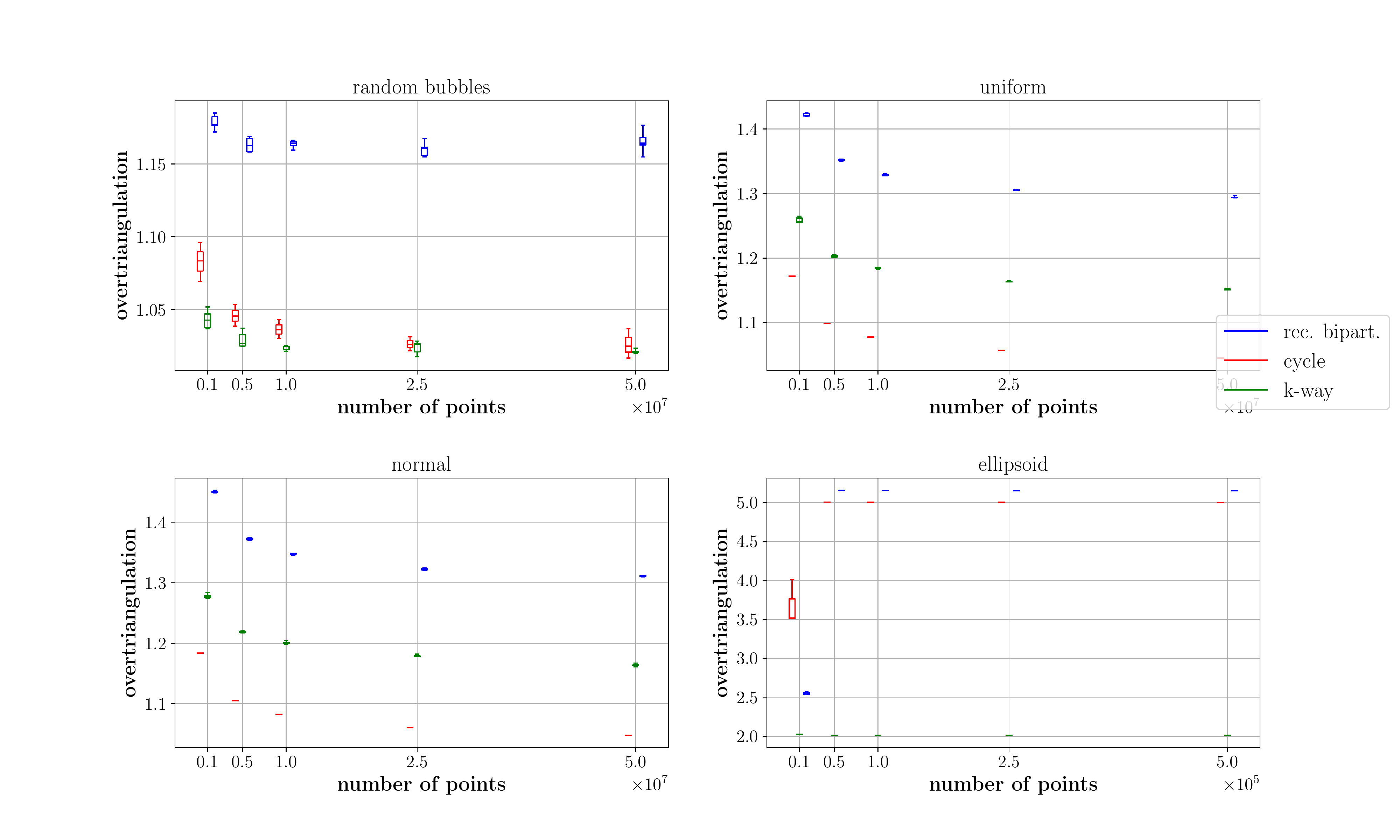}
    }{
        \begin{subfigure}{.49\textwidth}
        \centering
        \includegraphics[width=\textwidth,keepaspectratio]{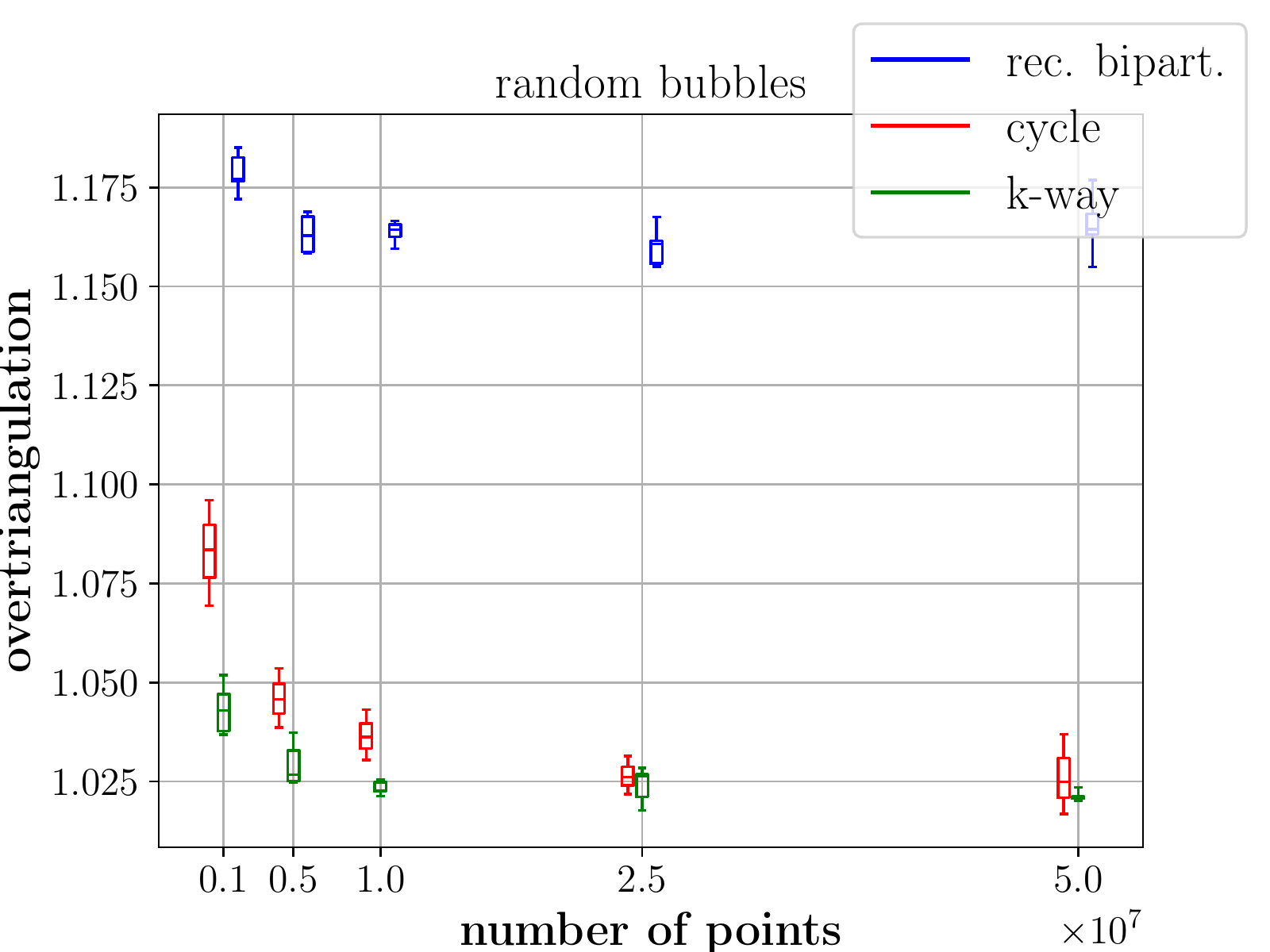}
        \end{subfigure}
        \begin{subfigure}{.49\textwidth}
        \centering
        \includegraphics[width=\textwidth,keepaspectratio]{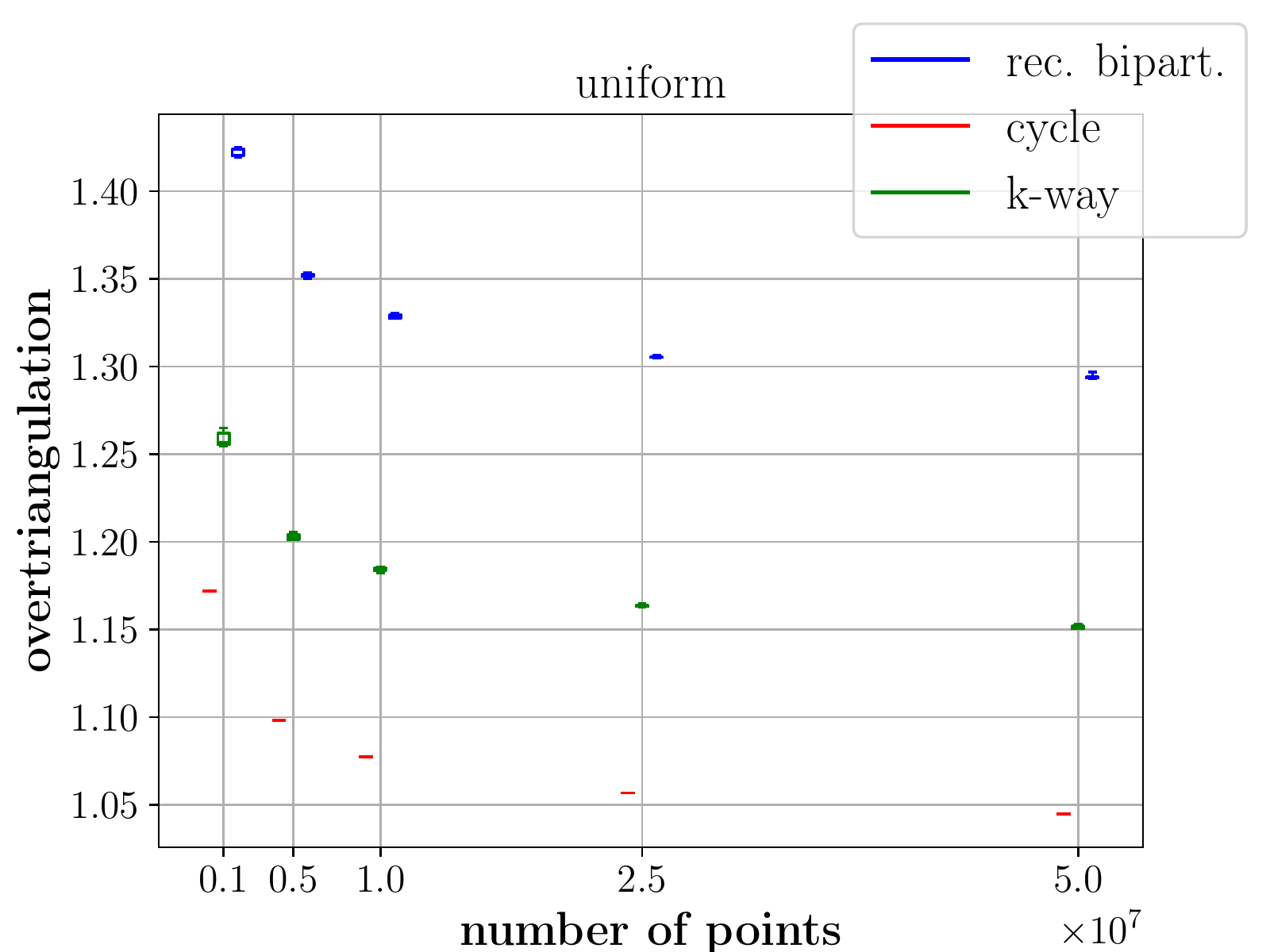}
        \end{subfigure}
    }
    \caption{sample size $\eta(n) = \num{.01}n$}
    \label{fig:ot:01}
    \end{subfigure}
    \begin{subfigure}{\textwidth}
    \centering
    \iftoggle{tr}{
        \includegraphics[width=.8\textwidth,keepaspectratio]{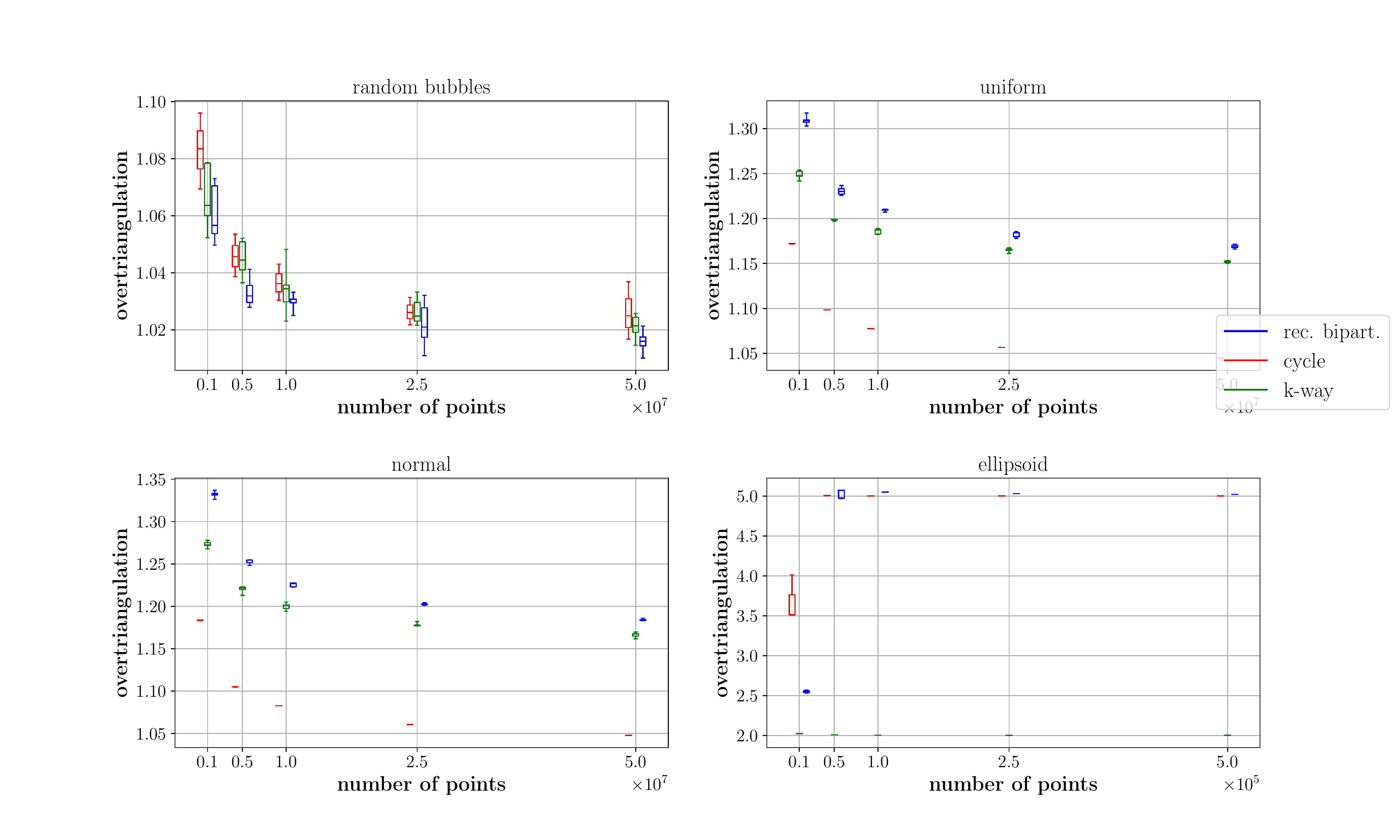}
    }{
        \begin{subfigure}{.49\textwidth}
        \centering
        \includegraphics[width=\textwidth,keepaspectratio]{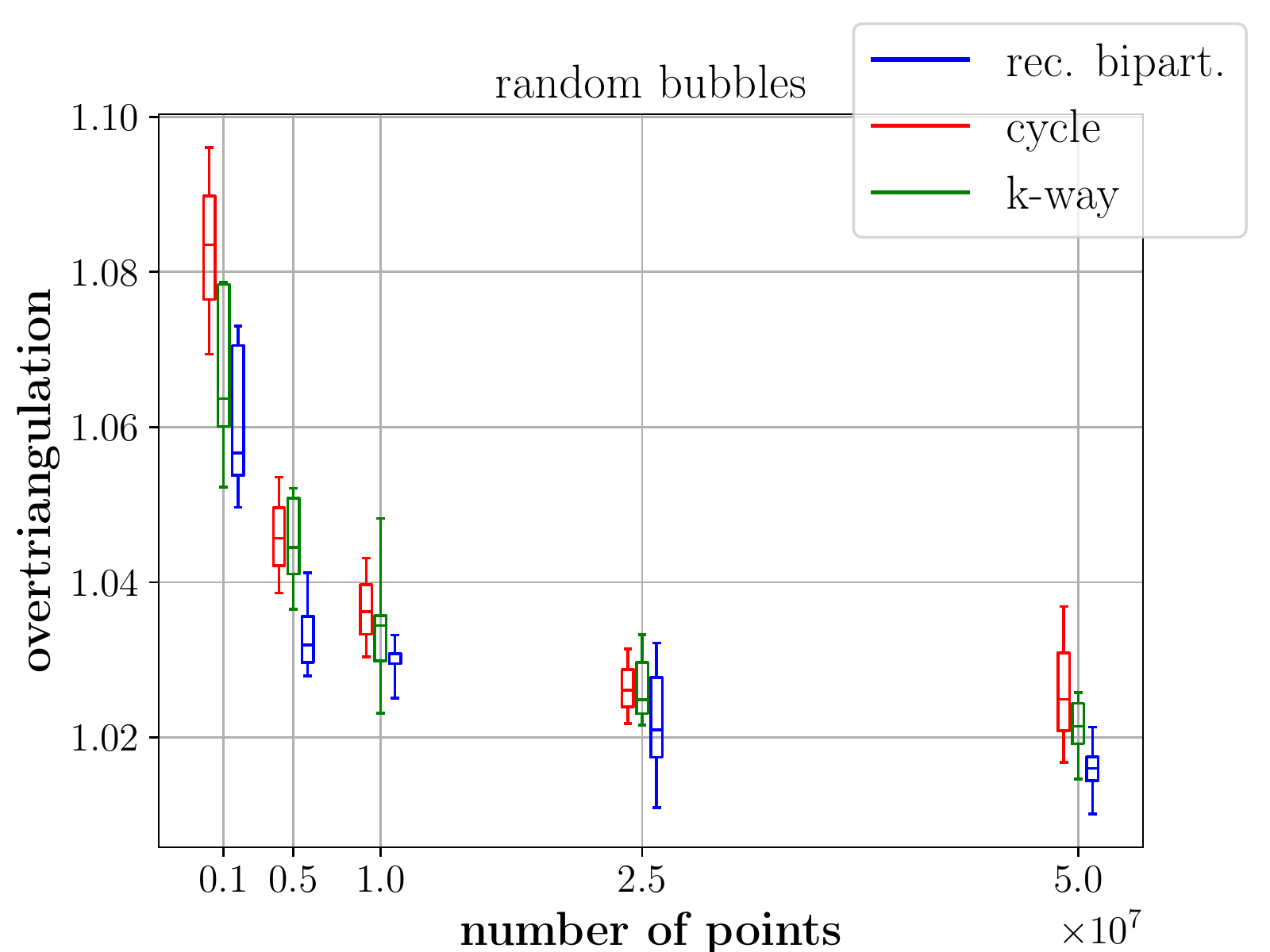}
        \end{subfigure}
        \begin{subfigure}{.49\textwidth}
        \centering
        \includegraphics[width=\textwidth,keepaspectratio]{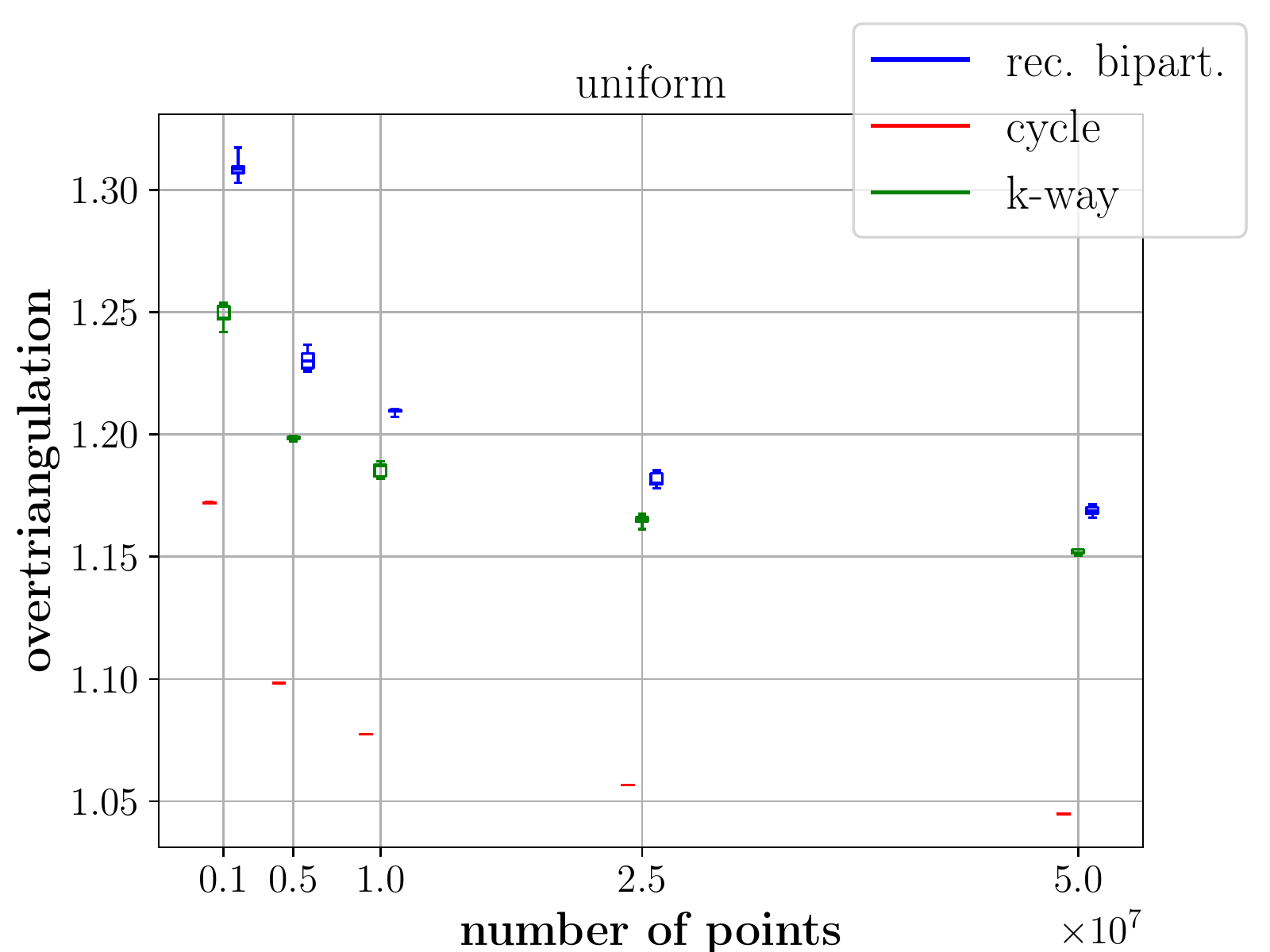}
        \end{subfigure}
    }
    \caption{sample size $\eta(n) = \sqrt{n}$}
    \label{fig:ot:sqrt}
    \end{subfigure}
   \caption{Overtriangulation factor for $k = t= 16$, parallel KaHIP, logarithmic edge weights and grid-based intersection test with $c_\mathcal{G} = 1$.}
  \label{fig:ot}
\end{figure}

Given a graph partitioning $\begin{pmatrix} V_1 & \dots & V_k \end{pmatrix}$,
its quality is defined by the weight of its cut,
 $\sum_{e \in C} \omega(e)$ for  $C := \{e = (u,v), e \in E \text{ and } u \in V_i, v \in V_j \text{ with } i \neq j\}$.
As mentioned in \secref{sec:sbp}, 
the balance of the graph partitioning is ensured by the imbalance parameter $\epsilon$,
$\left|V_i\right| \leq (1+\epsilon) \lceil  \frac{|V|}{k} \rceil$ for all $i \leq k$.
When the partitioning of the sample triangulation is extended to the entire input set,
this guarantee no longer holds.
We therefore study two quality measures:
\begin{enumerate*}[i)]
 \item the deviation from the ideal partition size and
 \item the coefficient of variation of the partition sizes.
\end{enumerate*}
\iftoggle{tr}{}{Due to space constraints we only discuss the latter measure here and refer to the technical report~\citep{fullpaper} for the former one.}

\iftoggle{tr}{
The deviation from the ideal partition size is given by $\nicefrac{p_i}{\frac{N}{k}} - 1$,
for $k$ partitions with $N$ points in total and partition sizes $p_i$, $i \leq k$,
and is shown in \figref{fig:idev} for a fixed choice of KaHIP configuration, edge weights and two different sample sizes.
Our sample-based approach produces almost equally sized partitions for the random bubble distribution
and clearly outperforms the cyclic partitioning scheme.
The larger sample size of $\num{.01}n$ results in a more balanced partitioning compared to $\sqrt{n}$.
Considering the uniform distribution, the cyclic partitioning scheme produces perfectly balanced partitions with smooth cuts between them,
whereas our new divide-step suffers from the jagged border between the partitions.
}{}

The coefficient of variation $c_v$ of the partition sizes \iftoggle{tr}{$p_i = |\set{P}_i|$}{$p_i$}, $i \leq k$, is
given by
\[
c_v = \frac{\sigma}{\mu} = \frac{\sqrt{\frac{\sum_{i \leq k} (p_i - \mu)^2}{k-1}}}{\frac{\sum_{i \leq k} p_i}{k}}.
\]
\figref{fig:dev} shows $c_v$ for\iftoggle{tr}{ a fixed choice of KaHIP configuration, edge weights and}{} two different sample sizes\iftoggle{tr}{}{
and two of our input distributions -- we refer to the technical report~\citep{fullpaper} for the other distributions}.
For all distributions, our sample-based partitioning scheme robustly achieves a $c_v$ of \SI{\approx 6}{\percent} and \SI{\approx 12}{\percent} for
sample sizes $\sqrt{n}$ and $\num{.01}n$, respectively.\iftoggle{tr}{\footnote{%
We attribute the outlier for the ellipsoid distribution to the small input size.}}{}
Both lie above the chosen imbalance of the graph partitioning of $\epsilon = \SI{5}{\percent}$, as expected.
The larger sample size not only decreases the average imbalance but also its spread for various random seeds.
Moreover, the deficits of the original cyclic partitioning scheme become apparent:
whereas it works exceptionally well for uniformly distributed points,
it produces inferior partitions in the presence of an underlying structure in the input,
as found for instance in the random bubble distribution.

In total, our recursive algorithm triangulates more than the number of input points due to
the triangulation of the sample points, and the triangulation(s) of the border point set(s).
We quantify this in the overtriangulation factor $o_{DT}$, given by
\[
o_{DT} := \frac{|\set{P}| + \sum |\set{P}_S| + \sum |\operatorname{vertices}(\set{B})|}{|\set{P}|}.
\]
$\set{B}$ is the set of border simplices\iftoggle{tr}%
{, refer to \lnref{ln:borderpoints} of \algref{alg:sma}}%
{ determined by our \dac algorithm}%
.
For direct $k$-way partitioning, only one sample and one border triangulation are necessary;
for recursive bisectitioning there are a total of $k-1$ of each.
\figref{fig:ot} shows the overtriangulation factor 
for\iftoggle{tr}{ a fixed choice of KaHIP configuration, edge weight and}{} two different sample sizes\iftoggle{tr}{}{
and two of our input distributions -- again, we refer to~\citep{fullpaper}}.
For all distributions, the larger sample size reduces the oversampling factor.
As the partitioning of the larger sample DT more closely resembles the partitioning of the full DT,
the number of points in the border triangulation is reduced.
For the random bubble distribution, the overtriangulation factor is on par or below that of
the original cyclic partitioning scheme.
\iftoggle{tr}{
The ellipsoid distribution is specifically tailored to be a hard input.
Due to its large convex hull, almost all points are part of the border triangulation,
therefore the oversampling factor is bound by the maximum recursion depth.
For the normally distributed input point set,
the central dense region needs to be cut multiple times in order to ensure balance between the partition size.
Thus, more points are part of the border point set.}{}
For the uniform distribution, our new divide-step suffers from the jagged border between the partitions 
compared to the smooth cut produced by the cyclic partitioning scheme.
This results in more circumhyperspheres intersecting another partition and thus the inclusion of more points in the border triangulation.
Our experiments with the exact intersection test primitive confirm this notion.

\subsection{Runtime Evaluation}\label{sec:eval:runtime}

\iftoggle{tr}{
\begin{figure}[tb]
	\centering
	\includegraphics[width=.8\textwidth,keepaspectratio]{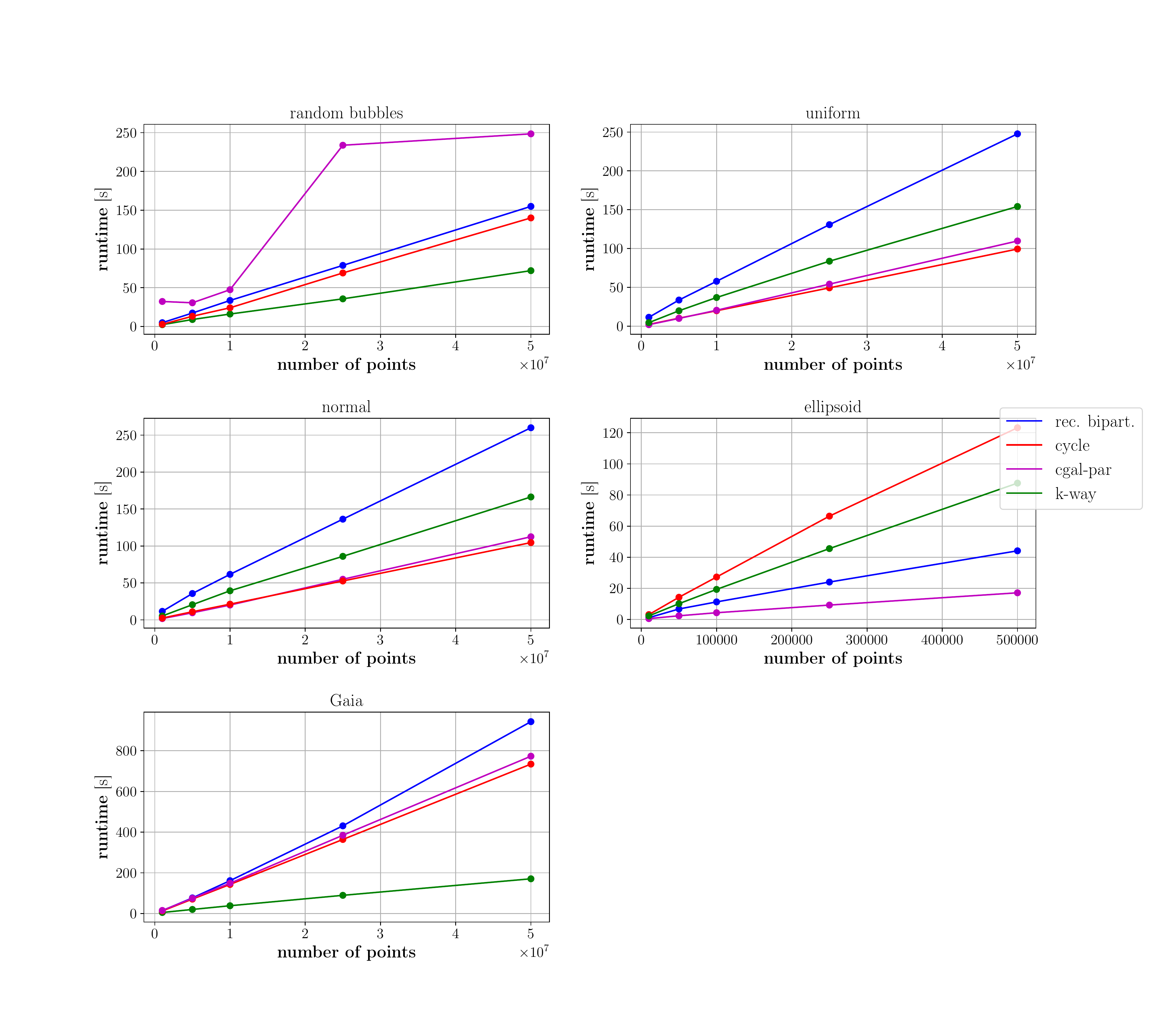}
	\caption{Runtime evaluation for $k = t= 16$, parallel KaHIP, $\eta(n) = \sqrt{n}$, grid-based intersection test with $c_\mathcal{G} = 1$ and logarithmic edge weights.}
	\label{fig:runtime}
\end{figure}
}{ 
\begin{figure}[tb]
	\centering
    \begin{subfigure}{.49\textwidth}
    \centering
    \includegraphics[width=\textwidth,keepaspectratio]{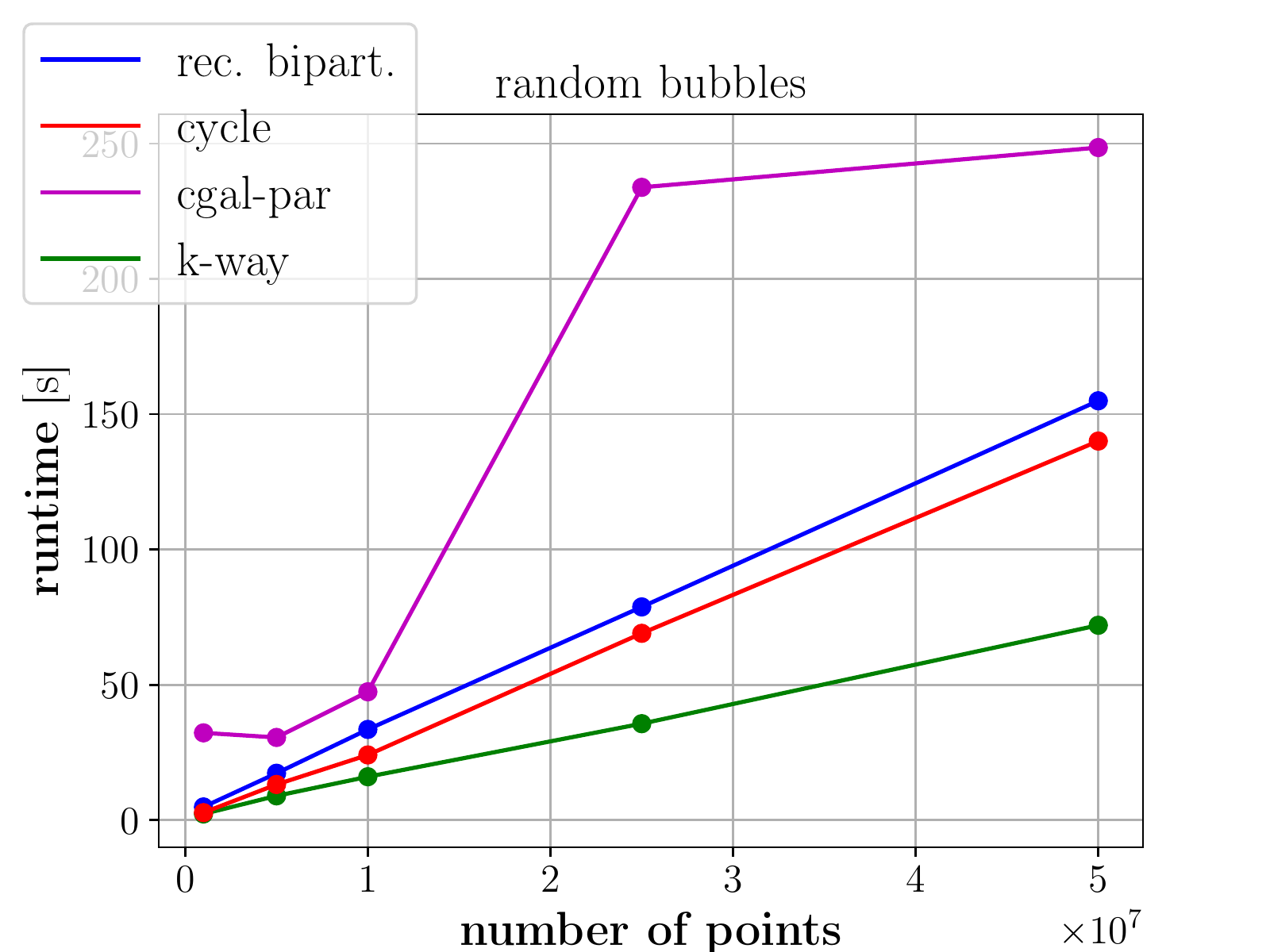}
    \caption{patch bubble}
    \label{fig:runtime:pb}
    \end{subfigure}
    \begin{subfigure}{.49\textwidth}
    \centering
    \includegraphics[width=\textwidth,keepaspectratio]{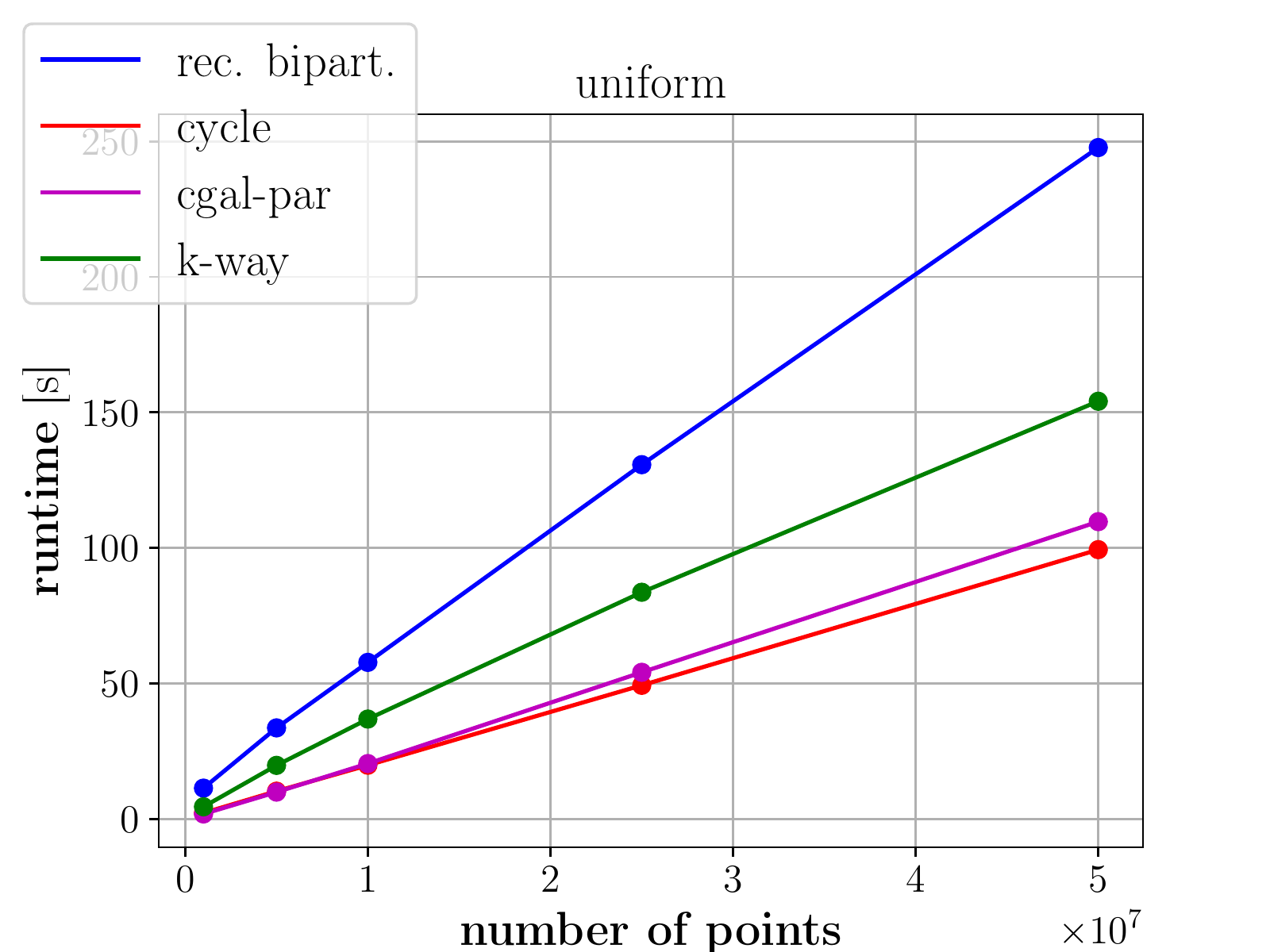}
    \caption{uniform}
    \label{fig:runtime:uni}
    \end{subfigure}
    \begin{subfigure}{.49\textwidth}
    \centering
    \includegraphics[width=\textwidth,keepaspectratio]{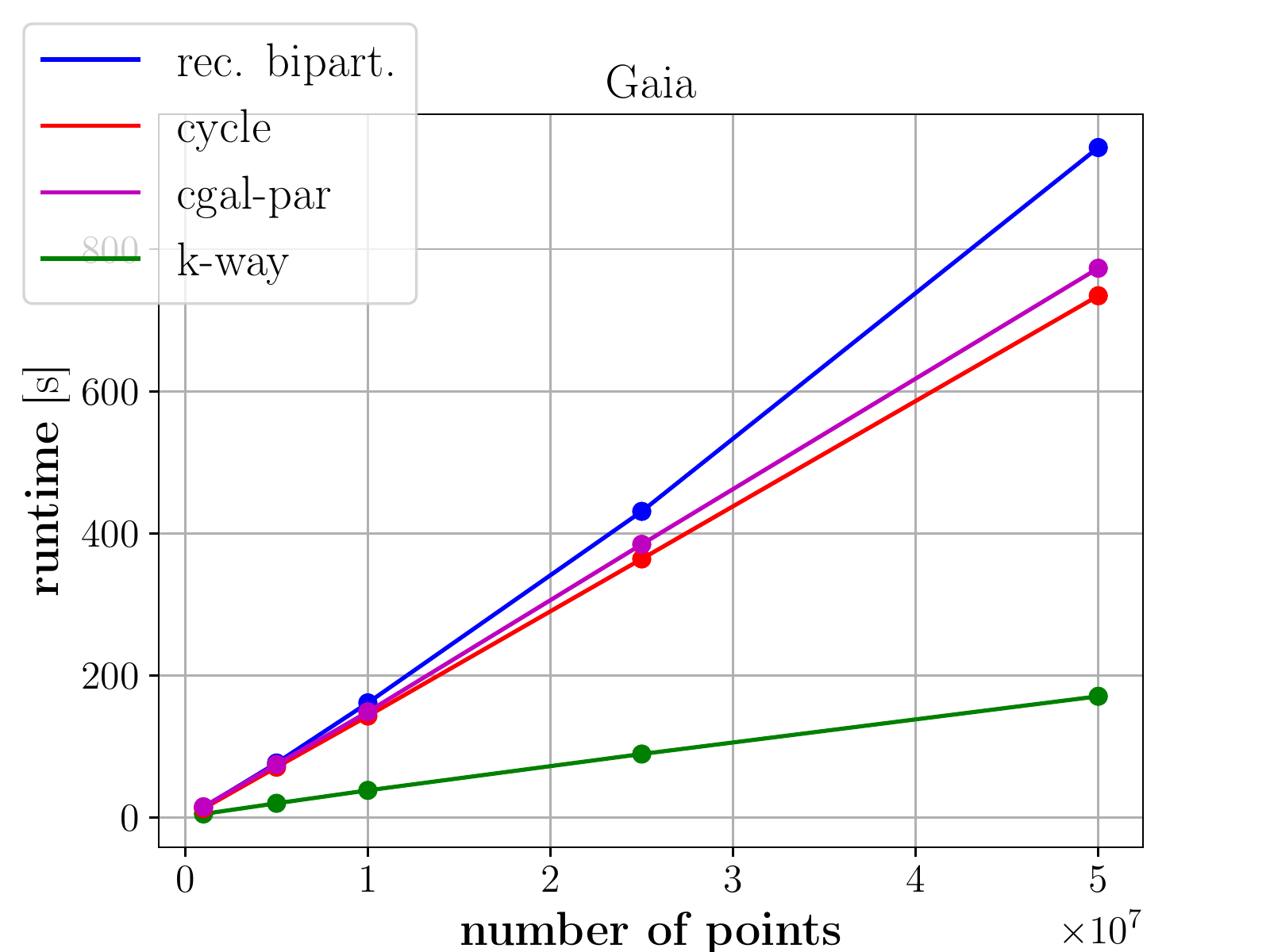}
    \caption{Gaia}
    \label{fig:runtime:gaia}
    \end{subfigure}
    \begin{subfigure}{.49\textwidth}
    \centering
    \includegraphics[width=\textwidth,keepaspectratio]{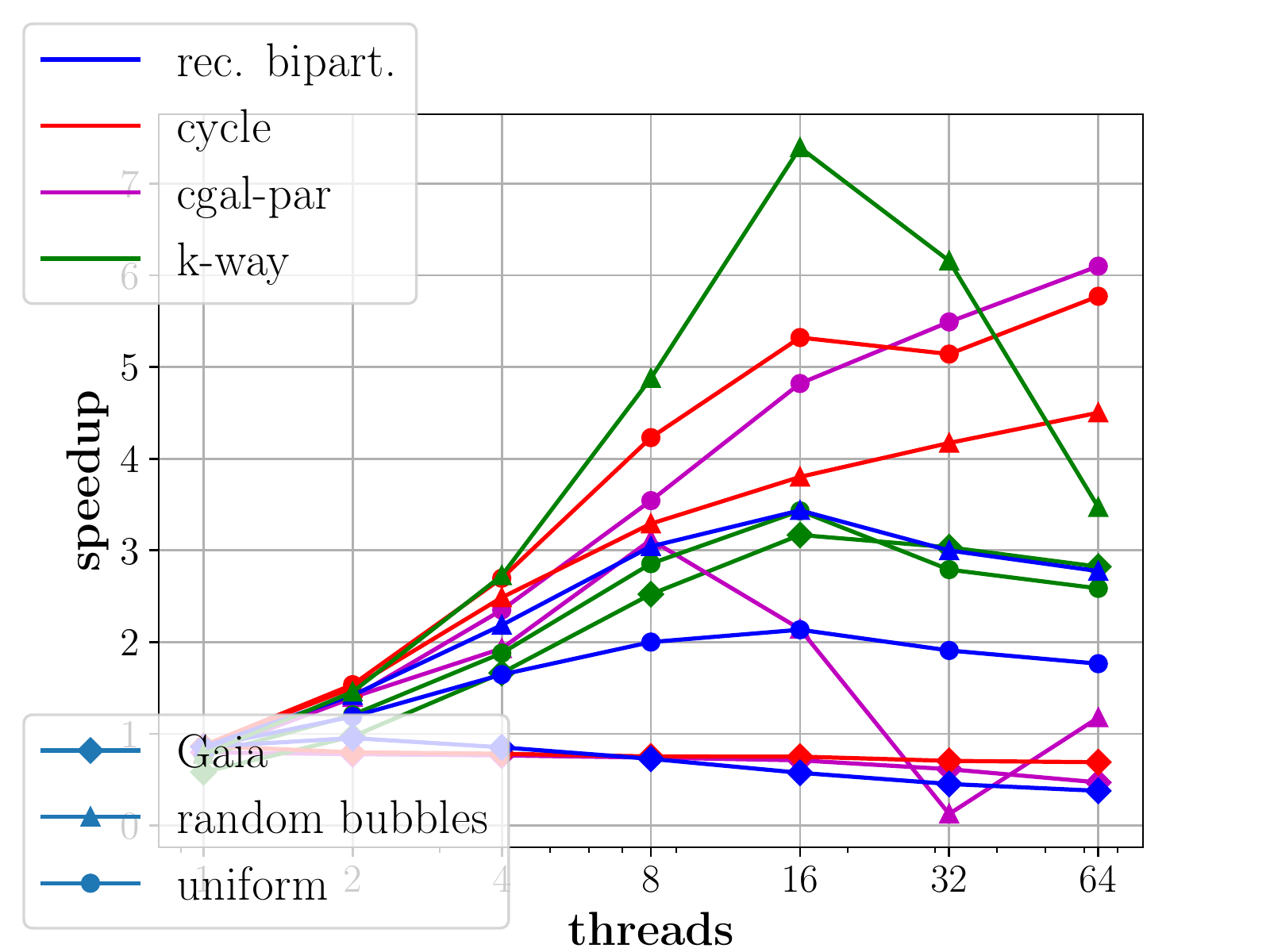}
    \caption{speedup}
    \label{fig:speedup}
    \end{subfigure}
   \caption{Runtime evaluation for $k = t= 16$, parallel KaHIP, $\eta(n) = \sqrt{n}$, grid-based intersection test with $c_\mathcal{G} = 1$ and logarithmic edge weights.
   Absolute speedup over sequential CGAL for $k = t$ and all distributions tested with \num{50e6} points.}
  \label{fig:runtime}
\end{figure}
}

\iftoggle{tr}{
\begin{figure}[tb]
  \centering
  \includegraphics[width=.7\textwidth,keepaspectratio]{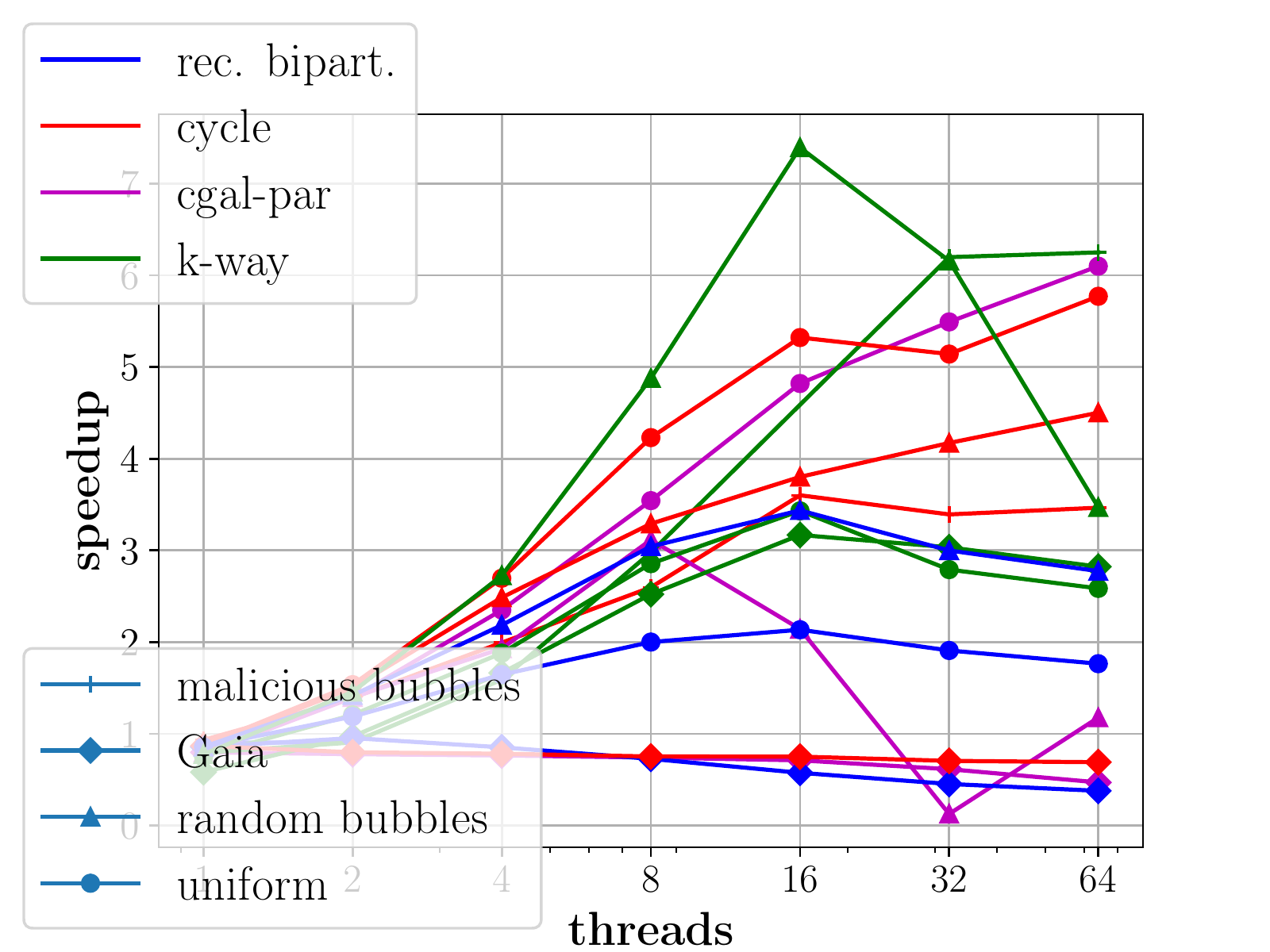}
  \caption{Absolute speedup over sequential CGAL for $k = t$, parallel KaHIP, $\eta(n) = \sqrt{n}$, grid-based intersection test with $c_\mathcal{G} = 1$ and logarithmic edge weights.
  All distributions are tested with \iftoggle{tr}{the maximum number of points given in \tabref{tab:pointsets}}{\num{50e6} points}.}
  \label{fig:speedup}
\end{figure}
}

We conclude our experiments with a study of the runtime of \iftoggle{tr}{\algref{alg:sma}}{our \dac algorithm} with our sample-based divide step against
the originally proposed cyclic division strategy as well as the parallel incremental insertion algorithm of CGAL.
\figref{fig:runtime} shows the total triangulation time for \iftoggle{tr}{a fixed choice of KaHIP configuration, edge weights and sample size}
{our fixed choice of configuration parameters}.


Direct $k$-way partitioning performs best on the random bubbles distribution,
with a speedup of up to \SI{50}{\percent} over the cyclic partitioning scheme.
CGAL's parallel incremental insertion algorithm requires locking to avoid race conditions.
It therefore suffers from high contention in the bubble centers,
resulting in a \SI{350}{\percent} speedup for our approach.
For uniformly distributed points, our new divide-step falls behind
the cyclic partitioning scheme as there is no structure to exploit in the input data 
and due to the higher overtriangulation factor, as discussed in the previous section.
However, comparing an $o_{DT} = 1.15$ for $k$-way partitioning to $o_{DT} = 1.05$ for cyclic partitioning
-- about a \SI{10}{\percent} increase --
only explains part of the \SI{52}{\percent} slowdown.
Further investigation is therefore required to identify -- and mitigate -- the source of the remaining slowdown.

Of particular interest is the scaling behavior of our algorithm with an increasing number of threads.
\figref{fig:speedup} shows a strong scaling experiment\iftoggle{tr}{ for a fixed choice of KaHIP configuration, edge weights and sample size}{}.
The absolute speedup over the sequential CGAL algorithm is given by $\text{Speedup}(t) := \frac{T_\text{CGAL}}{T(t)}$ for $t$ threads.


In the presence of exploitable input structure -- such as for the random bubble distribution --
direct $k$-way partitioning scales well on one physical processor (up to 16 cores).
It clearly outperforms the original cyclic partitioning scheme and the parallel DT algorithm of CGAL.
Nevertheless, it does not scale well to two sockets ($t > 16$ threads) and hyper-threading ($t > 32$ threads).
The overtriangulation factor of \num{1.19} for 64 threads compared to \num{1.015} for 16 suggests
that the size of the input is not sufficient to be efficiently split into 64 partitions.

Considering our real world dataset, the direct $k$-way partitioning scheme also exhibits the best scaling behavior.
\iftoggle{tr}{As illustrated in \figref{fig:gaia}, t}{T}he dataset comprises a large dense \iftoggle{tr}{ring}{area} accompanied by several smaller isolated regions.
This can be exploited to reduce border triangulation sizes and achieve a speedup,
compared to the slowdown for the cyclic partitioning scheme and CGAL's parallel algorithm.
The former is due to large border triangulations in the central ring, 
whereas the latter suffers from contention in the central region.

\iftoggle{tr}{
The performance for normally distributed points can be attributed to the high overtriangulation factor,
refer to\iftoggle{tr}{ \figref{fig:ot} and its discussion in}{} the previous section.}{}

Clearly, direct $k$-way partitioning outperforms recursive bisection in every configuration.
Following the theoretical considerations in \secref{sec:strategies} regarding the number of merge-steps required,
this is to be expected.
A measure to level the playing field would be to only
allow for $\eta(n)$ \emph{total} number of sample points on all levels,
\ie adjust the sample size on each level of the recursion according the expected halving of the input size.

\section{Conclusions}\label{sec:outro}

We present a novel divide-step for the parallel \dac DT algorithm presented in~\citep{Funke2017}.
The input is partitioned according to the graph partitioning of a Delaunay triangulation of a small input point sample.
The partitioning scheme robustly delivers well-balanced partitions for all tested input point distributions.
For input distributions exhibiting an exploitable underlying structure,
it further leads to small border triangulations and fast merging.
On favorable inputs, we achieve almost a factor of two speedup over our previous partitioning scheme
and over the parallel DT algorithm of CGAL.
These inputs include synthetically generated data sets as well as
the Gaia DR2 star catalog.
For uniformly distributed input points, the more complex divide-step incurs an overall runtime penalty compared to the original approach,
opening up two lanes of future work:
\begin{enumerate*}[i)]
 \item smoothing the border between the partitions to reduce the overtriangulation factor, and/or
 \item an adaptive strategy that chooses between the classical partitioning scheme and our new approach based on
 easily computed properties of the chosen sample point set, before computing its DT. 
\end{enumerate*}
\iftoggle{tr}{
Furthermore, building on the idea of~\citet{Lee2001},
the partition borders could be traced with Delaunay edges to avoid merging all together.}{}
The sample-based divide step can also be integrated into our distributed memory algorithm presented in~\citep{Funke2017},
where the improved load-balancing and border size reduces the required communication volume for favorable inputs.

\iftoggle{tr}{%
\bibliographystyle{plainnat}
\bibliography{misc,cgal,delaunay}
}{
\bibliographystyle{styles/splncs04nat}
\bibliography{extracted}

\begin{thebibliography}{32}
\providecommand{\natexlab}[1]{#1}
\providecommand{\url}[1]{\texttt{#1}}
\expandafter\ifx\csname urlstyle\endcsname\relax
  \providecommand{\doi}[1]{doi: #1}\else
  \providecommand{\doi}{doi: \begingroup \urlstyle{rm}\Url}\fi

\bibitem[Aggarwal et~al.(1988)Aggarwal, Chazelle, and Guibas]{Aggarwal1988}
A~Aggarwal, B~Chazelle, and L~Guibas.
\newblock {Parallel computational geometry}.
\newblock \emph{Algorithmica}, 3\penalty0 (1):\penalty0 293--327, 1988.

\bibitem[Akhremtsev et~al.(2017)Akhremtsev, Heuer, Sanders, and
  Schlag]{kahypar}
Y.~Akhremtsev, T.~Heuer, P.~Sanders, and S.~Schlag.
\newblock {Engineering a direct \emph{k}-way Hypergraph Partitioning
  Algorithm}.
\newblock In \emph{Workshop on Algorithm Engineering and Experiments,
  (ALENEX)}, pages 28--42. SIAM, 2017.

\bibitem[Akhremtsev et~al.(2018)Akhremtsev, Sanders, and Schulz]{parhip}
Yaroslav Akhremtsev, Peter Sanders, and Christian Schulz.
\newblock High-quality shared-memory graph partitioning.
\newblock In Marco Aldinucci, Luca Padovani, and Massimo Torquati, editors,
  \emph{Euro-Par 2018: Parallel Processing}, pages 659--671. Springer, 2018.

\bibitem[Batista et~al.(2010)Batista, Millman, Pion, and Singler]{Batista2010}
Vicente~H.F. Batista, David~L. Millman, Sylvain Pion, and Johannes Singler.
\newblock {Parallel geometric algorithms for multi-core computers}.
\newblock \emph{Computational Geometry}, 43\penalty0 (8):\penalty0 663--677,
  2010.

\bibitem[Bentley(1975)]{Bentley1975}
JL~Bentley.
\newblock {Multidimensional binary search trees used for associative
  searching}.
\newblock \emph{Communications of the ACM}, 18\penalty0 (9):\penalty0 509--517,
  1975.

\bibitem[Blelloch et~al.(1999)Blelloch, Miller, Hardwick, and
  Talmor]{Blelloch1999}
E.~G. Blelloch, L.~G. Miller, C.~J. Hardwick, and D.~Talmor.
\newblock Design and implementation of a practical parallel delaunay algorithm.
\newblock \emph{Algorithmica}, 24\penalty0 (3):\penalty0 243--269, 1999.

\bibitem[Chen(2010)]{Chen2010}
Min-Bin Chen.
\newblock {The Merge Phase of Parallel Divide-and-Conquer Scheme for 3D
  Delaunay Triangulation}.
\newblock In \emph{International Symposium on Parallel and Distributed
  Processing with Applications (ISPA)}, pages 224--230. IEEE, 2010.

\bibitem[Chen and Gotsman(2012)]{Chen2012}
R.~Chen and C.~Gotsman.
\newblock Localizing the delaunay triangulation and its parallel
  implementation.
\newblock In \emph{International Symposium on Voronoi Diagrams in Science and
  Engineering (ISVD)}, pages 24--31. IEEE, June 2012.

\bibitem[Cheng et~al.(2012)Cheng, Dey, and Shewchuk]{Cheng2012}
Siu-Wing Cheng, Tamal~K Dey, and Jonathan Shewchuk.
\newblock \emph{Delaunay mesh generation}.
\newblock CRC Press, 2012.

\bibitem[Chrisochoides(2006)]{Chrisochoides2006}
Nikos Chrisochoides.
\newblock Parallel mesh generation.
\newblock In Are~Magnus Bruaset and Aslak Tveito, editors, \emph{Numerical
  Solution of Partial Differential Equations on Parallel Computers}, pages
  237--264. Springer, 2006.

\bibitem[Chrisochoides and Nave(2000)]{Chrisochoides2000}
Nikos Chrisochoides and Démian Nave.
\newblock Simultaneous mesh generation and partitioning for delaunay meshes.
\newblock \emph{Mathematics and Computers in Simulation}, 54\penalty0
  (4):\penalty0 321 -- 339, 2000.

\bibitem[Cignoni et~al.(1998)Cignoni, Montani, and Scopigno]{Cignoni1998}
P~Cignoni, C~Montani, and R~Scopigno.
\newblock {DeWall: A fast divide and conquer Delaunay triangulation algorithm
  in $E^d$}.
\newblock \emph{Computer-Aided Design}, 30\penalty0 (5), 1998.

\bibitem[Collaboration(2018)]{gaiadr2}
Gaia Collaboration.
\newblock Gaia data release 2. summary of the contents and survey properties.
\newblock \emph{arXiv}, \penalty0 (abs/1804.09365), 2018.

\bibitem[Delaunay(1934)]{Delaunay1934}
B.~Delaunay.
\newblock {Sur la sph\`ere vide. A la m\'emoire de Georges Vorono\"\i}.
\newblock \emph{{Bulletin de l'Acad\'emie des Sciences de l'URSS. Classe des
  Sciences Math\'ematiques et Naturelles}}, \penalty0 (6):\penalty0 793--800,
  1934.

\bibitem[Devillers(2002)]{dt_hierarchy}
O.~Devillers.
\newblock The delaunay hierarchy.
\newblock \emph{International Journal of Foundations of Computer Science},
  13\penalty0 (02):\penalty0 163--180, 2002.

\bibitem[Frazer and McKellar(1970)]{samplesort}
W.~D. Frazer and A.~C. McKellar.
\newblock Samplesort: A sampling approach to minimal storage tree sorting.
\newblock volume~17, pages 496--507. ACM, July 1970.

\bibitem[Fuetterling et~al.(2014)Fuetterling, Lojewski, and
  Pfreundt]{Fuetterling2014}
Valentin Fuetterling, Carsten Lojewski, and Franz-Josef Pfreundt.
\newblock {High-Performance Delaunay Triangulation for Many-Core Computers}.
\newblock In \emph{Eurographics/ ACM SIGGRAPH Symposium on High Performance
  Graphics}. The Eurographics Association, 2014.

\bibitem[Funke and Sanders(2017)]{Funke2017}
D.~Funke and P.~Sanders.
\newblock Parallel $d$-d delaunay triangulations in shared and distributed
  memory.
\newblock In \emph{Workshop on Algorithm Engineering and Experiments (ALENEX)},
  pages 207--217. SIAM, 2017.

\bibitem[Hert and Seel(2015)]{cgal:hs-chdt3-15b}
Susan Hert and Michael Seel.
\newblock {dD} convex hulls and delaunay triangulations.
\newblock In \emph{{CGAL} User and Reference Manual}. {CGAL Editorial Board},
  {4.7} edition, 2015.

\bibitem[Karypis and Kumar(1998)]{metis}
G.~Karypis and V.~Kumar.
\newblock A fast and high quality multilevel scheme for partitioning irregular
  graphs.
\newblock \emph{SIAM Journal on Scientific Computing}, 20\penalty0
  (1):\penalty0 359--392, 1998.

\bibitem[Kernighan and Lin(1970)]{gpPractice}
B.~W. Kernighan and S.~Lin.
\newblock {An Efficient Heuristic Procedure for Partitioning Graphs}.
\newblock \emph{{The Bell System Technical Journal}}, 49\penalty0 (2):\penalty0
  291--307, Feb 1970.

\bibitem[Kohout et~al.(2005)Kohout, Kolingerov\'{a}, and
  \v{Z}\'{a}ra]{Kohout2005}
Josef Kohout, Ivana Kolingerov\'{a}, and Ji{\v r}\'{\i} \v{Z}\'{a}ra.
\newblock {Parallel Delaunay triangulation in E2 and E3 for computers with
  shared memory}.
\newblock \emph{Parallel Computing}, 31\penalty0 (5):\penalty0 491--522, 2005.

\bibitem[Larsson et~al.(2007)Larsson, Akenine-M\"{o}ller, and
  Lengyel]{circletest}
Thomas Larsson, Tomas Akenine-M\"{o}ller, and Eric Lengyel.
\newblock {On Faster Sphere-Box Overlap Testing}.
\newblock \emph{Journal of Graphics, GPU, and Game Tools}, 12\penalty0
  (1):\penalty0 3--8, 2007.

\bibitem[Lee et~al.(2001)Lee, Park, and Park]{Lee2001}
Sangyoon Lee, Chan-Ik Park, and Chan-Mo Park.
\newblock An improved parallel algorithm for delaunay triangulation on
  distributed memory parallel computers.
\newblock \emph{Parallel Processing Letters}, 11:\penalty0 341--352, 2001.

\bibitem[Sanders and Schulz(2011)]{kahip_params}
Peter Sanders and Christian Schulz.
\newblock Engineering multilevel graph partitioning algorithms.
\newblock In Camil Demetrescu and Magn{\'u}s~M. Halld{\'o}rsson, editors,
  \emph{Algorithms -- ESA 2011}, pages 469--480, Berlin, Heidelberg, 2011.
  Springer.

\bibitem[Sanders and Schulz(2013)]{kahip}
Peter Sanders and Christian Schulz.
\newblock {Think Locally, Act Globally: Highly Balanced Graph Partitioning}.
\newblock In \emph{Proc. of Int. Symp. on Experimental Algorithms (SEA'13)},
  volume 7933 of \emph{LNCS}, pages 164--175. Springer, 2013.

\bibitem[Sanders et~al.(2018)Sanders, Lamm, H\"{u}bschle-Schneider, Schrade,
  and Dachsbacher]{sampling}
Peter Sanders, Sebastian Lamm, Lorenz H\"{u}bschle-Schneider, Emanuel Schrade,
  and Carsten Dachsbacher.
\newblock Efficient parallel random sampling -- vectorized, cache-efficient,
  and online.
\newblock \emph{ACM Trans. Math. Softw.}, 44\penalty0 (3):\penalty0
  29:1--29:14, January 2018.

\bibitem[Shewchuk(1996)]{Shewchuk1996}
JR~Shewchuk.
\newblock {Triangle: Engineering a 2D quality mesh generator and Delaunay
  triangulator}.
\newblock \emph{Applied Computational Geometry Towards Geometric Engineering},
  1148:\penalty0 203--222, 1996.

\bibitem[Shewchuk(1997)]{Shewchuk1997}
JR~Shewchuk.
\newblock Adaptive {P}recision {F}loating-{P}oint {A}rithmetic and {F}ast
  {R}obust {G}eometric {P}redicates.
\newblock \emph{Discrete \& Computational Geometry}, 18\penalty0 (3):\penalty0
  305--363, October 1997.

\bibitem[Simon and Teng(1997)]{gpTheory}
H.~D. Simon and S.-H. Teng.
\newblock How good is recursive bisection?
\newblock \emph{SIAM Journal on Scientific Computing}, 18\penalty0
  (5):\penalty0 1436--1445, September 1997.

\bibitem[Su and Drysdale(1995)]{Su1995}
Peter Su and Robert L.~Scot Drysdale.
\newblock A comparison of sequential delaunay triangulation algorithms.
\newblock In \emph{Symposium on Computational Geometry (SCG)}, pages 61--70.
  ACM, 1995.

\bibitem[van~den Bergen(1997)]{aabb}
Gino van~den Bergen.
\newblock Efficient collision detection of complex deformable models using aabb
  trees.
\newblock \emph{Journal of Graphics Tools}, 2\penalty0 (4):\penalty0 1--13,
  1997.

\end{thebibliography}
}

\end{document}